\documentclass[trackchanges]{aastex701}

\usepackage{booktabs}
\usepackage{float}
\usepackage{placeins}

\begin{document}

\title{Dust Activity Evolution and Constraints on Icy Grains in the Interstellar Comet 3I/ATLAS}

\author{Bin liu}
\affiliation{State Key Laboratory of Lunar and Planetary Sciences, Macau University of Science and Technology, Taipa, Macau 999078, People's Republic of China}
\email{liub97@126.com}

\author{Xiaoping Zhang}
\affiliation{State Key Laboratory of Lunar and Planetary Sciences, Macau University of Science and Technology, Taipa, Macau 999078, People's Republic of China}
\email{xpzhangnju@gmail.com}

\author{Xiaodong Liu}
\affiliation{School of Aeronautics and Astronautics, Shenzhen Campus of Sun Yat-sen University, Shenzhen, Guangdong 518107, People's Republic of China}
\email{liuxd36@mail.sysu.edu.cn}

\author{Man-To Hui}
\affiliation{Shanghai Astronomical Observatory, Chinese Academy of Sciences, No. 80 Rd Nandan, Shanghai 200030, Mainland China}
\email{manto@hawaii.edu}

\correspondingauthor{Xiaoping Zhang}

\email{xpzhangnju@gmail.com}

\begin{abstract}
We present Zwicky Transient Facility observations of the interstellar comet 3I/ATLAS to characterize its dust-activity evolution before and after perihelion. Before perihelion, the effective scattering cross section $C_{\rm e}$ brightened across the inner $2\times10^4$ km as $C_{\rm e}\propto r_{\rm H}^{-\gamma}$, with $\gamma$ increasing from $1.63\pm0.14$ to $1.88\pm0.34$ with aperture size. After perihelion, the heliocentric index decreased from $2.44\pm0.09$ in the $5\times10^3$ km aperture to $1.54\pm0.09$ in the $8\times10^4$ km aperture, indicating faster fading of the inner coma. Under adopted grain-size and residence-time assumptions, the implied dust mass-loss rates are $\sim123$--$203~{\rm kg~s^{-1}}$ before perihelion and $\sim596$--$1412~{\rm kg~s^{-1}}$ after perihelion. The morphology yields a sunward ejection-speed scale of $v_{\rm ej}\sim(190\pm3)a_{\mu{\rm m}}^{-1/2}~{\rm m~s^{-1}}$ and an effective perpendicular velocity of $v_\perp=(79\pm6)a_{\mu{\rm m}}^{-1/2}~{\rm m~s^{-1}}$. Post-perihelion surface-brightness profiles transiently steepened to $k\sim-2.7$ in late November and early December, then became shallower and approached $k=-1.5$ by January. Matched $g-r$ profiles show that this phase coincided with an anomalously blue outer coma that disappeared by January. We characterize the fading component with a projected destruction length of $\ell_{\rm DE}\sim(6.4$--$7.2)\times10^4$ km. Thermophysical sublimation calculations indicate that this scale, together with the order of magnitude of the water production near the main imaging epoch, is consistent with $a_0\sim5$--$10~\mu{\rm m}$ icy grains containing a small absorbing carbonaceous fraction of $f_{\rm C}\sim0.1\%$. These results suggest that the early post-perihelion coma likely contained a short-lived, volatile-bearing icy-grain component whose fading contributed to the outer-coma brightness and color evolution.
\end{abstract}



\section{Introduction} 
Interstellar objects are rare samples of planetesimal material that formed outside the Solar System and was later ejected into interstellar space \citep{kareta2025near}. The first two confirmed cases, 1I/'Oumuamua and 2I/Borisov, revealed the diversity among such bodies, with 1I/'Oumuamua showing no unambiguous cometary coma, while 2I/Borisov displayed activity more similar to that of Solar System comets \citep{hoogendam2025university}. Comet 3I/ATLAS, hereafter 3I, was identified as the third confirmed interstellar object from its strongly hyperbolic orbit, with \(e \sim 6.1\) and \(V_\infty \sim 58~{\rm km~s^{-1}}\), and is notable for detected dust activity on the inbound leg at \(r_{\rm H} \sim 6.5~{\rm au}\) \citep{rahatgaonkar2025very,maggiolo2026interstellar,ye2025prediscovery}. As an active interstellar comet, 3I offers a rare opportunity to examine extrasolar cometary material through its observable activity. These observations can help constrain the properties and activity of material formed in another planetary system and provide a basis for comparison with comets formed in the Solar System \citep{rahatgaonkar2025very}.

The morphology of comet 3I's coma reflects the spatial distribution of dust released by ongoing activity and can provide constraints on the properties and evolution of the ejecta. Early imaging revealed an unusual morphology dominated by a narrow sunward dust structure rather than by a simple antisolar tail \citep{jewitt2025hubble,chandler2026nsf}.  This feature has been attributed to anisotropic dust release from a locally heated active region on a slowly rotating nucleus \citep{serra2025pre}, although projection effects involving icy grains with direction-dependent survival times have also been proposed \citep{keto2026physics}. These interpretations suggest that the observed coma cannot be explained by a steady, isotropic dust outflow alone. The activity level also increased rapidly as 3I approached perihelion \citep{zhang2026rapid}. Published measurements show \(Af\rho\) increasing from a few \(\times10^2\) cm shortly after discovery to \(\sim 2\times10^4\) cm near perihelion \citep{chandler2026nsf,ren2026interstellar}. Model-dependent estimates suggest a corresponding rise in the dust mass-loss rate from tens of kg s\(^{-1}\) to order \(10^3\) kg s\(^{-1}\) \citep{jewitt2025hubble,ren2026interstellar}. The pre- and post-perihelion evolution appears asymmetric, which may indicate changes in the active source regions, thermal processing of the surface, or evolving observing geometry \citep{roth2026isotopic,hui2026nucleus}. 

Spectroscopic observations provide complementary constraints on the composition of the coma and support the presence of icy material mixed with refractory dust. Optical and near-infrared spectra show a red continuum similar to D-type asteroid material, together with a broad absorption near \(2.0~\mu{\rm m}\) that has been interpreted as water ice in the coma \citep{yang2025spectroscopic}. Spectral-mixture models can reproduce the continuum and absorption feature with an ice component at the level of several tens of percent \citep{yang2025spectroscopic,keto2026physics}. Independent infrared observations also show water-ice absorption and an extended CO\(_2\) coma \citep{lisse2026spherex}.

Polarimetric observations provide additional constraints on the dust properties. The unusually deep and narrow negative polarization branch of comet 3I suggests an atypical scattering medium, possibly involving micron-sized icy grains mixed with dark refractory material \citep{gray2025extreme,zubko2025unusual}. Post-perihelion polarimetry has yielded a complex picture, with \citet{zheltobryukhov2025aperture} reporting an apparent change in the polarization phase curve inside the water snow line, which they associated with the loss of the water-ice component, whereas \citet{choi2026dust} found the $R_C$-band polarization phase curve to remain continuous across perihelion and instead attributed the unusual polarization primarily to refractory dust. Their photopolarimetric measurements sampled cometocentric radii of approximately $(1.5$--$2.0)\times10^{4}$~km, and therefore primarily constrain the scattering properties of the inner coma. Recent calculations further show that the survival of micron-sized icy grains on the outbound leg depends sensitively on grain size and optical properties \citep{zubko2026lifetime}. Thus, the available polarimetric measurements do not uniquely determine whether a weaker volatile-bearing grain population may persist on larger coma scales.

Previous studies have characterized the activity evolution of 3I using observations obtained over limited orbital intervals or measurements assembled from multiple facilities. These studies have revealed rapid changes in dust activity and possible asymmetry before and after perihelion. However, comparisons among heterogeneous data sets may be affected by differences in photometric calibration, aperture definition, observing geometry, and phase correction. A homogeneous photometric baseline across perihelion is therefore needed to better isolate the temporal evolution of the coma from cross-instrument systematics. Such a baseline also allows possible links between large-scale coma morphology and the properties of the released material to be examined more directly. In particular, the time-dependent surface-brightness and color structure of the coma may retain information on the projected distribution of the ejecta and on changes in the relative contribution of icy and refractory grains. If icy grains fade or are destroyed after release, their changing contribution may be reflected in both the radial brightness profile and the radial color gradient of the coma. These radial signatures can then be interpreted in terms of the post-release evolution of volatile-bearing grains. For grains with different sizes and carbonaceous fractions, equilibrium temperatures determine sublimation lifetimes and survival distances. Comparing these survival distances with the observed scales of brightness fading and color change provides complementary photometric and thermophysical constraints on the grain properties. This approach complements spectroscopic and polarimetric studies by testing whether a transient volatile-bearing grain component can survive over the observed coma scales and reproduce the radial fading and color changes.

In this work, we use ZTF observations to examine the activity evolution of 3I and to constrain the properties of volatile-bearing grains in its coma. Sect.~\ref{Data} presents the archival ZTF observations used in this study. Sect.~\ref{Results} describes the photometric measurements and morphological analysis. In Sect.~\ref{Dicussion}, we first quantify the heliocentric dependence of the activity, estimate the velocity component of the ejecta perpendicular to the orbital plane, and derive the corresponding dust mass-loss rates. We then examine departures from a radiation-pressure-modified surface-brightness profile and interpret the observed brightness and color gradients using destruction lengths set by the sublimation and outward transport of icy grains. The inferred icy-grain population is further assessed by comparing its predicted water production rate with the observed value. Sect.~\ref{Conclusions} summarizes the main conclusions.
\begin{table}
\caption{Observation Log}
\label{tab:obs_log}
\footnotesize
\setlength{\tabcolsep}{8pt}

\begin{center}

\begin{tabular}{lccccccccc}
  \toprule
Date & $r_{\rm H}^{a}$ & $\Delta^{b}$ & $\alpha^{c}$ & PsAng$^{d}$ & PsAMV$^{e}$ & $\delta^{f}$ & Scale & FWHM$_{\rm PSF}$ & FWHM$_{\rm coma}$\\
(UT) & (AU)            & (AU)         & (deg)        & (deg)   & (deg)& (deg)    & (km pixel$^{-1}$) & (pixel)& (pixel)\\
  \midrule
  29 Jul 2025 05:03 & 3.58 & 2.85 & 12.7 & 100.091 &  98.577 & -0.49 & 2093 & 2.29 & 3.81\\
  30 Jul 2025 05:04 & 3.55 & 2.84 & 13.1 & 100.123 &  98.742 & -0.47 & 2082 & 2.24 & 3.59\\
  31 Jul 2025 05:03 & 3.52 & 2.82 & 13.5 & 100.164 &  98.908 & -0.44 & 2071 & 1.99 & 3.51\\
  31 Jul 2025 05:07 & 3.52 & 2.82 & 13.5 & 100.165 &  98.908 & -0.44 & 2071 & 2.15 & 3.55\\
  9 Aug 2025 05:06  & 3.23 & 2.71 & 17.0 & 100.812 & 100.412 & -0.19 & 1991 & 2.00 & 3.79\\
  11 Aug 2025 05:53 & 3.16 & 2.69 & 17.7 & 101.001 & 100.749 & -0.13 & 1977 & 2.17 & 4.05\\
  12 Aug 2025 05:28 & 3.13 & 2.68 & 18.1 & 101.094 & 100.911 & -0.10 & 1970 & 2.27 & 4.46\\
  13 Aug 2025 05:03 & 3.10 & 2.68 & 18.4 & 101.190 & 101.072 & -0.07 & 1964 & 1.69 & 3.62\\
  21 Aug 2025 04:37 & 2.85 & 2.62 & 20.8 & 101.979 & 102.337 &  0.18 & 1923 & 2.32 & 4.22\\
  25 Nov 2025 12:07 & 1.68 & 1.98 & 30.0 & 293.928 & 109.931 &  2.38 & 1450 & 2.10 & 9.06\\
  28 Nov 2025 13:15 & 1.74 & 1.94 & 30.5 & 293.785 & 109.949 &  2.39 & 1424 & 2.41 & 9.24\\
  29 Nov 2025 11:38 & 1.77 & 1.93 & 30.6 & 293.742 & 109.947 &  2.39 & 1416 & 3.12 & 9.34\\
  6 Dec 2025 12:40  & 1.94 & 1.86 & 30.0 & 293.395 & 109.794 &  2.38 & 1363 & 1.91 & 9.32\\
  6 Dec 2025 13:12  & 1.94 & 1.86 & 30.0 & 293.394 & 109.793 &  2.38 & 1363 & 1.86 & 9.28\\
  8 Dec 2025 11:55  & 1.99 & 1.84 & 29.5 & 293.278 & 109.697 &  2.37 & 1352 & 1.70 & 9.16\\
  13 Dec 2025 11:57 & 2.12 & 1.81 & 27.5 & 292.917 & 109.320 &  2.30 & 1329 & 2.95 & 9.25\\
  16 Dec 2025 11:32 & 2.21 & 1.80 & 26.0 & 292.650 & 108.995 &  2.24 & 1322 & 2.47 & 9.28\\
  18 Dec 2025 09:56 & 2.26 & 1.80 & 24.8 & 292.456 & 108.740 &  2.19 & 1320 & 2.88 & 9.40\\
  20 Dec 2025 12:33 & 2.32 & 1.80 & 23.5 & 292.227 & 108.422 &  2.13 & 1320 & 2.57 & 9.42\\
  28 Dec 2025 10:23 & 2.56 & 1.84 & 17.7 & 291.270 & 106.856 &  1.85 & 1348 & 2.78 & 9.96\\
  12 Jan 2026 06:47 & 3.02 & 2.08 &  6.4 & 291.572 & 102.742 &  1.16 & 1525 & 2.73 & 9.46\\
  12 Jan 2026 10:06 & 3.02 & 2.08 &  6.3 & 291.631 & 102.700 &  1.15 & 1527 & 3.11 & 9.54\\
  14 Jan 2026 08:15 & 3.09 & 2.13 &  5.0 & 292.795 & 102.115 &  1.06 & 1562 & 2.19 & 9.12\\
  16 Jan 2026 07:06 & 3.15 & 2.18 &  3.8 & 295.015 & 101.520 &  0.97 & 1599 & 1.85 & 9.04\\
  18 Jan 2026 07:23 & 3.21 & 2.24 &  2.5 & 299.799 & 100.909 &  0.88 & 1641 & 1.94 & 7.72\\
  \bottomrule
\end{tabular}
\end{center}

\vspace{-1mm}
\noindent\footnotesize\textit{Notes.}\ 
$^{a}$ Heliocentric distance.
$^{b}$ Geocentric distance.
$^{c}$ Phase angle.
$^{d}$ Position angle of the extended Sun--target vector.
$^{e}$ Position angle of the negative heliocentric velocity vector.
$^{f}$ Angle between the observer and the target orbital plane.
\end{table}

\section{Data} \label{Data}
\begin{figure*}[htbp]
    \centering

    \begin{minipage}[t]{0.41\textwidth}
        \vspace{0pt}
        \centering
        \includegraphics[width=\linewidth]{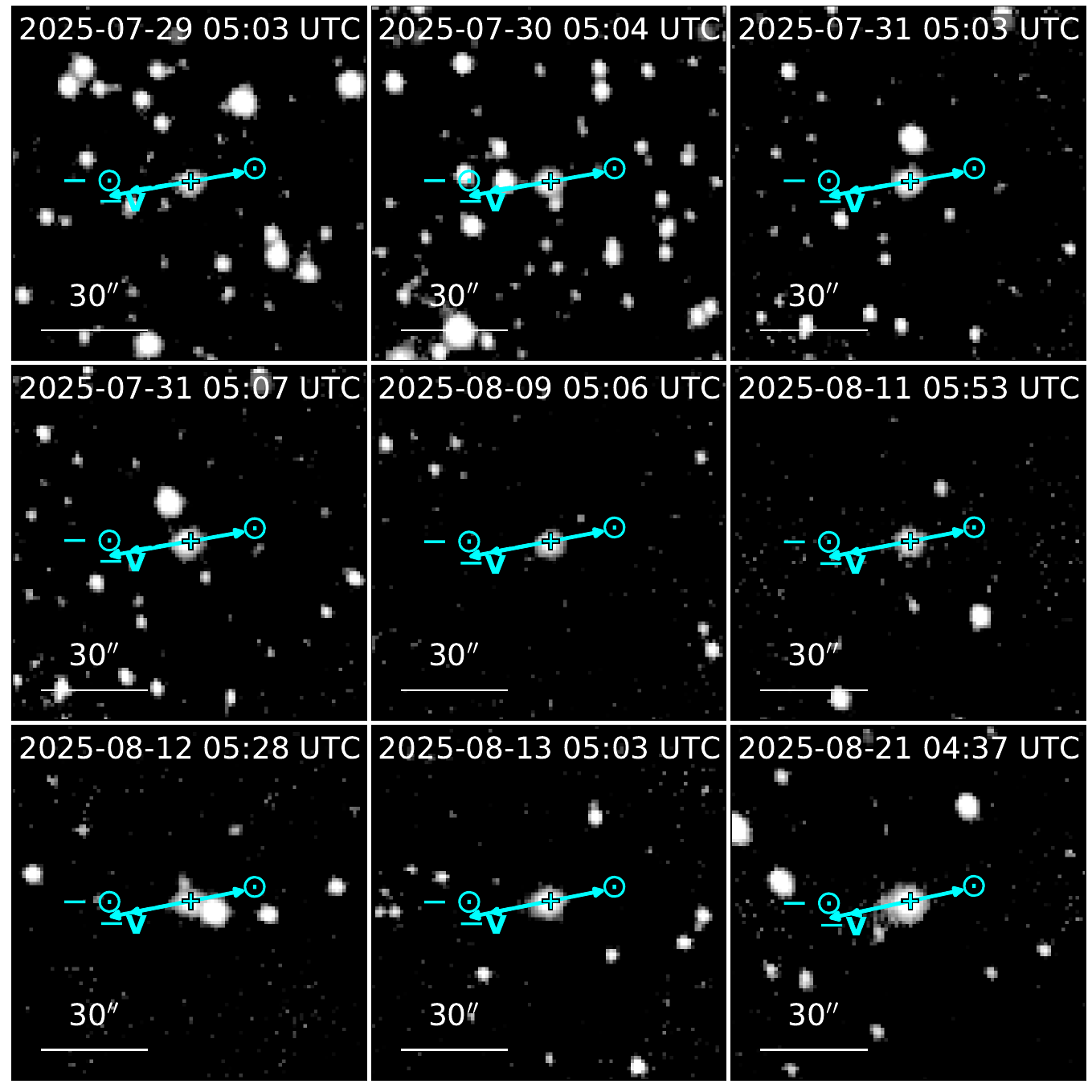}
        
        \vspace{0mm}
        {\small (a) Pre-perihelion images.}
    \end{minipage}
    \hfill
    \begin{minipage}[t]{0.58\textwidth}
        \vspace{0pt}
        \centering
        \includegraphics[width=\linewidth]{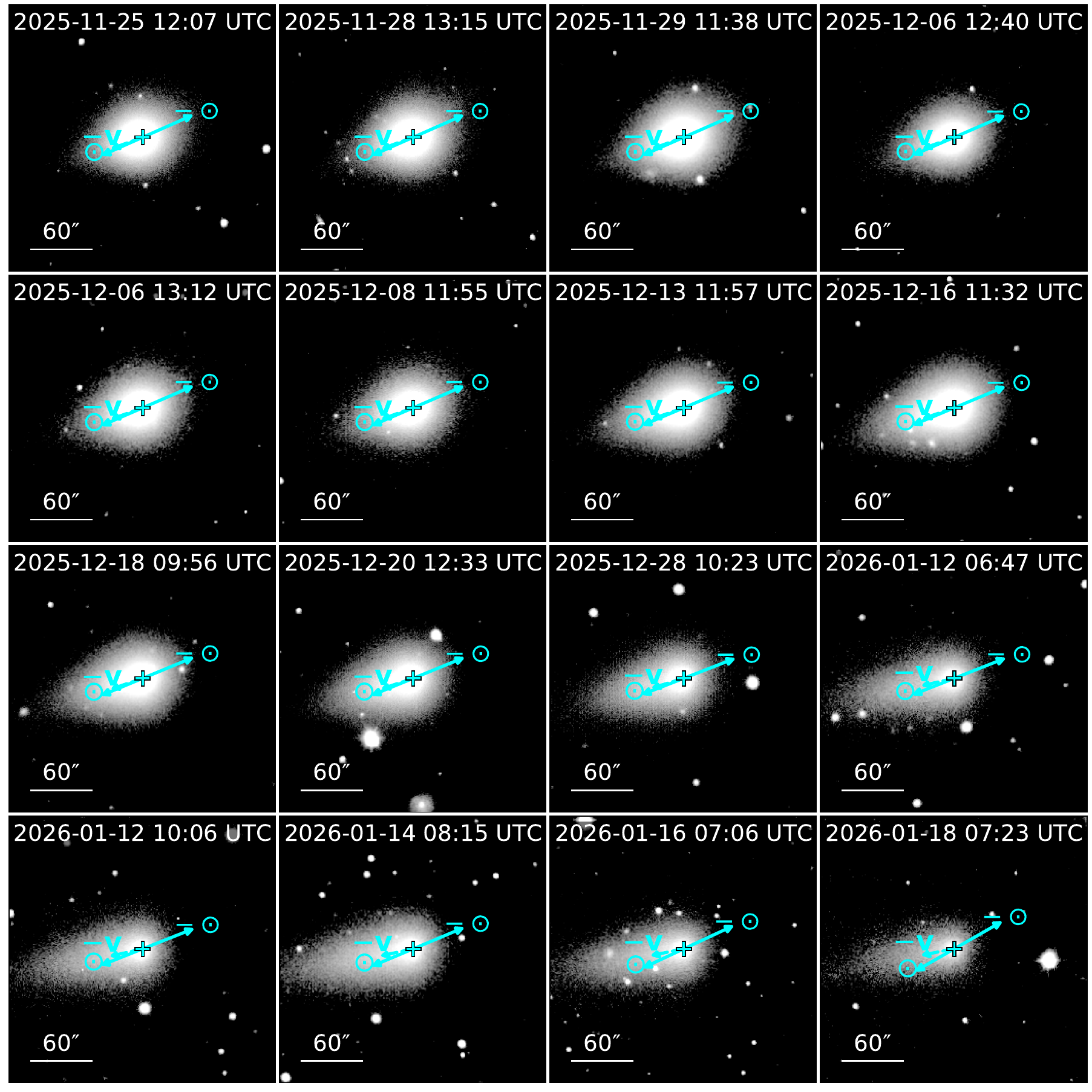}
        
        \vspace{0mm}
        {\small (b) Post-perihelion images.}
    \end{minipage}

    \caption{ZTF image sequences of 3I before and after perihelion. Panel (a) shows the pre-perihelion images, and panel (b) shows the post-perihelion images. The observation time is marked in each panel. The white scale bars indicate angular distances of 30$^{\prime\prime}$ in panel (a) and 60$^{\prime\prime}$ in panel (b). All images are oriented with north up and east to the left. The cyan arrows labeled $\odot$, $-\odot$, and $-\mathbf{v}$ indicate the projected solar, anti-solar, and anti-velocity directions, respectively, and the cyan cross marks the photometric center.}
    \label{fig:ztf_sequence}
\end{figure*}

We used images of comet 3I retrieved from the ZTF archive. The observations were obtained with the ZTF \(r\)-band filter, with exposure times of 30 s. The ZTF camera has a pixel scale of 1.012 arcsec pixel$^{-1}$. Images affected by poor image quality and low signal-to-noise ratio were excluded from the analysis. The observing geometries of the selected images, together with the measured stellar and comet FWHM values, are summarized in Tab.~\ref{tab:obs_log}, and the corresponding image sequences are shown in Fig.~\ref{fig:ztf_sequence}. The stellar PSF FWHM was measured from nearby field stars at each epoch. The comet FWHM is larger than the stellar PSF FWHM at every epoch. The data set was divided into two intervals corresponding to the pre- and post-perihelion phases. For each exposure, we extracted a square cutout centered on the photometric center of the comet. For the purpose of displaying the image sequence, a local background level was estimated from the median value of each cutout and subtracted. The images were then normalized by the background noise level and displayed using a common nonlinear stretch to enhance faint extended structures. 

Fig.~\ref{fig:ztf_sequence} shows the temporal evolution of the coma morphology. The observation time is marked in each panel, and the angular scale is indicated by the white scale bar. The pre-perihelion images show a relatively weak and compact source, whereas the post-perihelion images show a more extended coma and a tail-like structure. These images provide the visual context for the morphological and photometric analyses presented in the following sections.

\section{Results} \label{Results}
\subsection{Photometry} \label{Photometry}
To account for variations in atmospheric seeing, we convolved all ZTF images to a common PSF FWHM of 3.12 pixels, corresponding to the largest stellar FWHM in the sample. Remeasurements of field stars in the PSF-matched images confirmed that the final FWHM values were consistent with 3.12 pixels within the measurement uncertainties. We then performed fixed linear-aperture photometry on the PSF-matched images. The apertures were defined by projected physical radius and centered on the photometric center of the comet. Unlike fixed angular apertures, fixed linear apertures sample the same projected physical scale around the nucleus at different geocentric distances \(\Delta\), making measurements obtained at different epochs more directly comparable. We adopted projected radii of $5\times10^3$, $1\times10^4$, $2\times10^4$, $4\times10^4$, and $8\times10^4$~km. Before perihelion, the coma was only weakly extended, and its radial surface-brightness profile approached the background level beyond $\sim2\times10^4$~km. We therefore used only the three smallest apertures, $5\times10^3$, $1\times10^4$, and $2\times10^4$~km, for the pre-perihelion data. The two larger apertures, $4\times10^4$ and $8\times10^4$~km, were applied only after perihelion, when the coma had expanded sufficiently for the outer regions to remain measurable above the background.

Images affected by field-star contamination were first processed with a star-removal procedure, and epochs for which the contamination could not be corrected to a satisfactory level were excluded from further analysis. The total flux was then measured by summing the pixel values within each fixed aperture. The sky background was estimated from a concentric annulus placed outside the largest available photometric aperture, with inner and outer radii of approximately $(3$–$4)\times10^4$~km before perihelion and $(1$–$1.2)\times10^5$~km after perihelion. These ranges were chosen from epoch-stacked radial profiles so that the annulus lay in a region where the surface brightness had flattened to the background level and the residual coma contribution was small. Pixels in the annulus were sigma-clipped, and the median of the remaining pixels was adopted as the per-pixel sky background. This background level was scaled by the aperture area and subtracted from the aperture counts. The background-subtracted counts were converted to apparent \(r\)-band magnitudes using the photometric zero point of each image. The corresponding magnitude uncertainties were estimated from aperture counting noise, local sky variance, detector read noise, and the zero-point uncertainty.

The measured \(m_r\) values were corrected for heliocentric distance, geocentric distance, and phase angle to obtain absolute magnitudes,
\begin{equation}
H_r = m_r - 5\log_{10}(r_{\rm H}\Delta) - f (\alpha),
\label{eq:abs_mag}
\end{equation}
where $r_{\rm H}$ and $\Delta$ are the heliocentric and geocentric distances in au, respectively, $\alpha$ is the phase angle in degrees, and $f(\alpha)$ is the phase function. Here, we employed a linear phase function to correct for the phase angle effects. A linear phase function, \(f(\alpha)=\zeta\alpha\), was adopted. We used the phase coefficient \(\zeta = 0.026 \pm 0.006~{\rm mag~deg}^{-1}\) determined by \citet{hui2026nucleus}. This coefficient represents the intrinsic linear phase slope after accounting for the opposition surge at very small phase angles. The effective scattering cross section of the dust was then calculated from \(H_r\) as
\begin{equation}
C_{\rm e} = \frac{2.24 \times 10^{16}\pi}{p_r} \, 10^{0.4(m_{\odot,r} - H_r)},
\label{eq:cross_section}
\end{equation}
where \(C_{\rm e}\) is in km\(^2\), \(p_r=0.04\) is the assumed \(r\)-band geometric albedo of the dust, consistent with values commonly adopted for 3I \citep{ye2025prediscovery,jewitt2025preperihelion}, and \(m_{\odot,r}=-26.9\) is the adopted solar magnitude in the \(r\) band \citep{willmer2018absolute}. The resulting \(m_r\), \(H_r\), and \(C_{\rm e}\) values are listed in Appendix~\ref{appendix_auxiliary_figures} and Table~\ref{tab:cross}. The uncertainties were propagated from the magnitude errors and the adopted phase coefficient and were used in the subsequent analysis.

\begin{table*}
\centering
\renewcommand{\arraystretch}{1.15}
\caption{Effective scattering cross-sections measured in the ZTF $r$ band.}
\label{tab:cross}
\setlength{\tabcolsep}{6pt}

\textbf{Panel A. Pre-perihelion photometry}

\vspace{3pt}
\begin{tabular}{lccc}
    \toprule
    & \multicolumn{3}{c}{\(C_{\rm e}\) (km\(^{2}\))} \\
    \cmidrule(lr){2-4}
    UT Date
    & \(\rho=5\times10^{3}\) km
    & \(\rho=10\times10^{3}\) km
    & \(\rho=20\times10^{3}\) km \\
    \midrule
    29 Jul 2025 05:03 & 446 $\pm$ 32 & 729 $\pm$ 53 & 815 $\pm$ 61 \\
    30 Jul 2025 05:04 & 467 $\pm$ 35 & 723 $\pm$ 54 & 997 $\pm$ 61 \\
    31 Jul 2025 05:03 & 495 $\pm$ 38 & 804 $\pm$ 62 & 967 $\pm$ 74 \\
    31 Jul 2025 05:07 & 490 $\pm$ 38 & 789 $\pm$ 61 & 953 $\pm$ 74 \\
    09 Aug 2025 05:06 & 565 $\pm$ 56 & 964 $\pm$ 96 & 1165 $\pm$ 118 \\
    11 Aug 2025 05:53 & 573 $\pm$ 58 & 988 $\pm$ 101 & 1180 $\pm$ 122 \\
    12 Aug 2025 05:28 & 567 $\pm$ 60 & 838 $\pm$ 102 & 1065 $\pm$ 128 \\
    13 Aug 2025 05:03 & 584 $\pm$ 60 & 988 $\pm$ 102 & 1220 $\pm$ 126 \\
    21 Aug 2025 04:37 & 656 $\pm$ 77 & 1119 $\pm$ 131 & 1407 $\pm$ 165 \\
    \bottomrule
\end{tabular}

\vspace{8pt}

\textbf{Panel B. Post-perihelion photometry}

\vspace{3pt}
\begin{tabular}{lccccc}
    \toprule
    & \multicolumn{5}{c}{\(C_{\rm e}\) (km\(^{2}\))} \\
    \cmidrule(lr){2-6}
    UT Date
    & \(\rho=5\times10^{3}\) km
    & \(\rho=10\times10^{3}\) km
    & \(\rho=20\times10^{3}\) km
    & \(\rho=40\times10^{3}\) km
    & \(\rho=80\times10^{3}\) km \\
    \midrule
    25 Nov 2025 12:07 & 4039 $\pm$ 677 & 9213 $\pm$ 1545 & 17355 $\pm$ 2911 & 28454 $\pm$ 4772 & 40653 $\pm$ 6818 \\
    28 Nov 2025 13:15 & 3734 $\pm$ 636 & 8599 $\pm$ 1465 & 16496 $\pm$ 2811 & 27582 $\pm$ 4700 & 39812 $\pm$ 6784 \\
    29 Nov 2025 11:38 & 3521 $\pm$ 604 & 8258 $\pm$ 1416 & 16049 $\pm$ 2752 & 26981 $\pm$ 4627 & 39531 $\pm$ 6779 \\
    06 Dec 2025 12:40 & 2908 $\pm$ 488 & 6863 $\pm$ 1152 & 13730 $\pm$ 2304 & 24045 $\pm$ 4035 & 36023 $\pm$ 6045 \\
    06 Dec 2025 13:12 & 2993 $\pm$ 503 & 7001 $\pm$ 1176 & 13968 $\pm$ 2347 & 24541 $\pm$ 4123 & 36649 $\pm$ 6157 \\
    08 Dec 2025 11:55 & 2790 $\pm$ 461 & 6554 $\pm$ 1082 & 13153 $\pm$ 2171 & 23351 $\pm$ 3854 & 35067 $\pm$ 5788 \\
    13 Dec 2025 11:57 & 2381 $\pm$ 370 & 5630 $\pm$ 874 & 11499 $\pm$ 1785 & 21033 $\pm$ 3264 & 32914 $\pm$ 5108 \\
    16 Dec 2025 11:32 & 2140 $\pm$ 312 & 5097 $\pm$ 742 & 10451 $\pm$ 1522 & 19465 $\pm$ 2834 & 31349 $\pm$ 4565 \\
    18 Dec 2025 09:56 & 1982 $\pm$ 277 & 4711 $\pm$ 658 & 9700 $\pm$ 1354 & 18245 $\pm$ 2547 & 29598 $\pm$ 4133 \\
    20 Dec 2025 12:33 & 1843 $\pm$ 243 & 4441 $\pm$ 587 & 9206 $\pm$ 1216 & 17485 $\pm$ 2309 & 32600 $\pm$ 4306 \\
    28 Dec 2025 10:23 & 1369 $\pm$ 141 & 3527 $\pm$ 363 & 7568 $\pm$ 778 & 14446 $\pm$ 1485 & 24832 $\pm$ 2552 \\
    12 Jan 2026 06:47 & 854 $\pm$ 40 & 2183 $\pm$ 102 & 4747 $\pm$ 221 & 9686 $\pm$ 451 & 19226 $\pm$ 896 \\
    12 Jan 2026 10:06 & 924 $\pm$ 41 & 2232 $\pm$ 100 & 4765 $\pm$ 213 & 9612 $\pm$ 429 & 17764 $\pm$ 795 \\
    14 Jan 2026 08:15 & 910 $\pm$ 32 & 2162 $\pm$ 76 & 4578 $\pm$ 161 & 9323 $\pm$ 329 & 17432 $\pm$ 615 \\
    16 Jan 2026 07:06 & 893 $\pm$ 28 & 2102 $\pm$ 64 & 4424 $\pm$ 134 & 8957 $\pm$ 271 & 18213 $\pm$ 549 \\
    18 Jan 2026 07:23 & 860 $\pm$ 31 & 2053 $\pm$ 69 & 4316 $\pm$ 142 & 8755 $\pm$ 288 & 16007 $\pm$ 541 \\
    \bottomrule
\end{tabular}

\vspace{2pt}
\begin{flushleft}
\footnotesize
\textit{Note.} Here, \(\rho\) is the projected radius of the circular photometric aperture at the comet. The listed values are the effective scattering cross-sections \(C_{\rm e}\), in km\(^{2}\), with their associated uncertainties. The corresponding apparent and absolute magnitudes are provided in Appendix~\ref{appendix_auxiliary_figures}.
\end{flushleft}

\end{table*}

Before perihelion, the photometry shows a gradual brightening of the coma in all three apertures. In the 5000 km aperture, the apparent magnitude changed from \(m_r=17.47\pm0.02\) on 2025 July 29 to \(m_r=16.57\pm0.03\) on 2025 August 21. Over the same interval, the absolute magnitude changed from \(H_r=12.09\pm0.08\) to \(11.67\pm0.13\), and the effective scattering cross section increased from \(C_{\rm e}=(4.46\pm0.32)\times10^{2}\) to \((6.56\pm0.77)\times10^{2}~{\rm km}^{2}\). Similar brightening is seen in the 10000 km and 20000 km apertures (Tab.~\ref{tab:cross}). After perihelion, the trend reversed. Between 2025 November 25 and 2026 January 18, the 5000 km aperture faded from \(m_r=13.08\pm0.03\) to \(15.72\pm0.04\), while \(H_r\) increased from \(9.70\pm0.18\) to \(11.38\pm0.04\) and \(C_{\rm e}\) decreased from \((4.04\pm0.68)\times10^{3}\) to \((8.60\pm0.31)\times10^{2}~{\rm km}^{2}\). In the 80000 km aperture, which was measurable only after perihelion, \(m_r\) faded from \(10.57\pm0.03\) to \(12.55\pm0.03\), \(H_r\) increased from \(7.19\pm0.18\) to \(8.20\pm0.04\), and \(C_{\rm e}\) decreased from \((4.07\pm0.68)\times10^{4}\) to \((1.60\pm0.05)\times10^{4}~{\rm km}^{2}\).

\subsection{Morphology} \label{Morphology}
We used the pre- and post-perihelion images for two distinct but complementary morphological analyses. The pre-perihelion observations were used for the turnaround-distance analysis because the inbound coma was relatively compact and its sunward boundary was more clearly defined. In addition, the pre-perihelion morphology provides a cleaner approximation to the radiation-pressure turnaround picture, consistent with previous observations of a Sun-facing dust component and independent estimates of the dust ejection speed from the sunward coma extent \citep{jewitt2025hubble,ren2026interstellar}. By contrast, our post-perihelion profiles show evidence for grain fading or destruction, which can modify a brightness-defined extent without directly tracing the dynamical turnaround. We therefore regard the pre-perihelion measurements as the more direct proxy for the sunward turnaround scale. For each pre-perihelion image, we determined the projected sunward direction at the corresponding observing epoch and extracted the surface-brightness profile through the photometric center of the comet. Following the approach of \citet{jewitt2026investigation}, we characterized the sunward extent from the decline of the background-subtracted brightness profile. We adopted the same threshold of 10\% of the peak intensity and estimated the projected sunward turnaround distance, $R'_{\rm ta}$, as the distance from the photometric center of the comet to the first sunward position at which this threshold was reached.

Assuming that the measured sunward feature is approximately aligned with the Sun--comet direction, we converted the projected distance to the deprojected turnaround distance using $R_{\rm ta}=R'_{\rm ta}/\sin\alpha$, where $\alpha$ is the phase angle at the corresponding observing epoch. For dust initially ejected toward the Sun and subsequently decelerated by solar radiation pressure, the turnaround distance is approximated by
\begin{equation}
R_{\rm ta}=\frac{v_{\mathrm{ej}}^2 }{2 \beta g_{\text{sun}}},
	\label{vej}
\end{equation}
where $v_{\rm ej}$ is the ejection speed of the grains, $\beta$ is the ratio of radiation pressure to solar gravity. The local solar gravitational acceleration is given by $g_{\rm sun} = g_{\rm sun} {(1)}/r_{\rm H}^2$, where $g_{\rm sun} {(1)}=0.006$ m/s$^2$ is its value at heliocentric distance $r_{\rm H}=1$ au. For each image, we calculated the velocity coefficient $v_{0,i}=(2g_{{\rm sun},i}R_{{\rm ta},i})^{1/2}$ from the measured
sunward extent. The projected sunward extents and the corresponding ejection-speed estimates, together with their uncertainties, are listed in Table~\ref{tab:sunward_extent}. The nine pre-perihelion measurements yield a median velocity coefficient of $v_0=189.9$~m~s$^{-1}$, with an epoch-to-epoch median absolute
deviation of $3.0$~m~s$^{-1}$. We therefore adopt $v_{\rm ej}=(190\pm3)\beta^{1/2}~{\rm m~s^{-1}}$ for the following analysis. Because this deprojection assumes that the sunward turnaround lies along the Sun--comet axis, the derived velocity should be interpreted as an effective sunward ejection-speed scale.

\begin{figure*}
\centering
    \includegraphics[width=0.8\columnwidth]{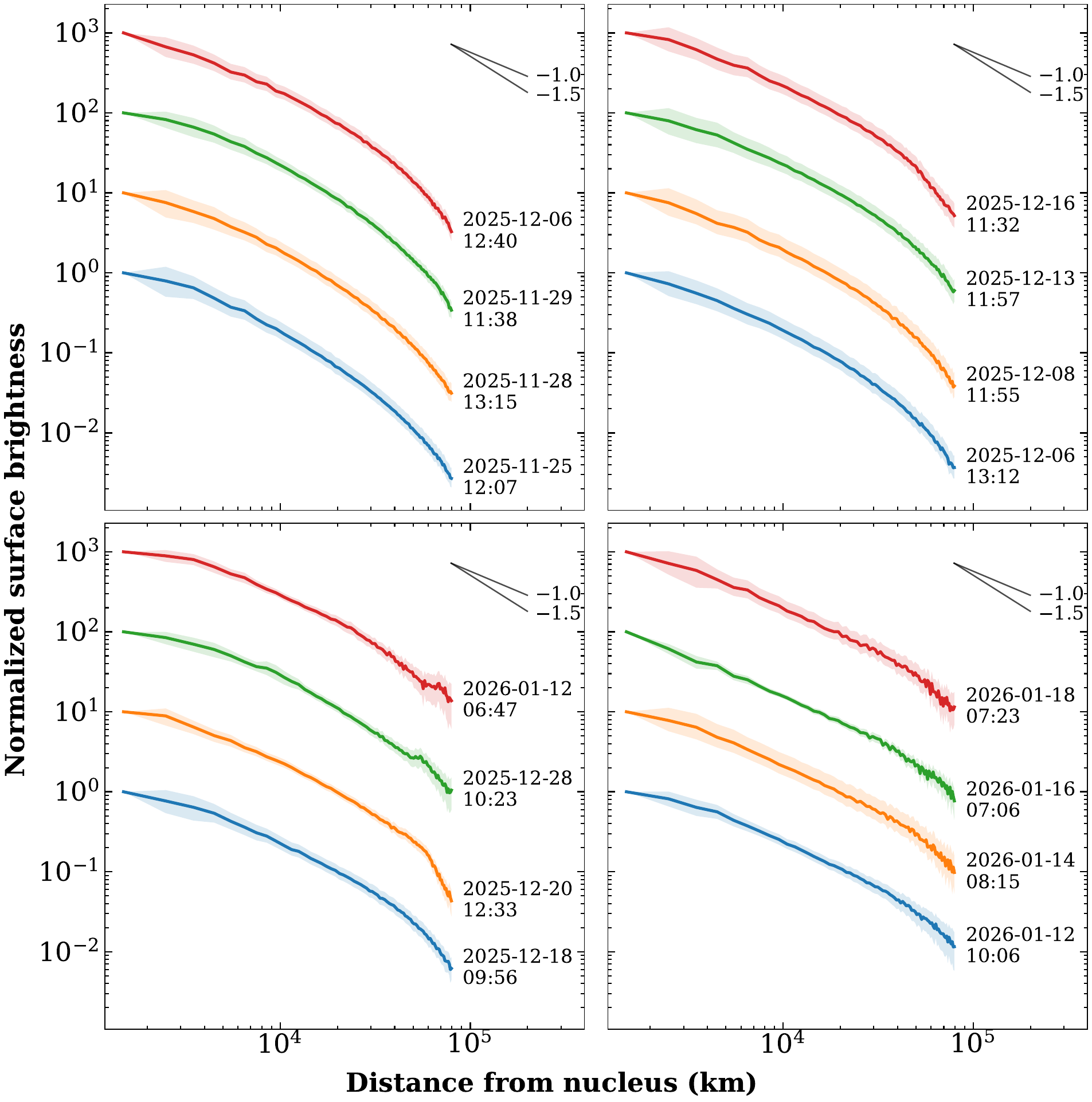}
    \caption{Normalized radial surface brightness profiles of the comet at 16 epochs. The profiles are shown in log-log space and grouped into four panels. Within each panel, small vertical offsets are applied for clarity. The shaded regions show the surface-brightness uncertainties. Short reference segments with slopes of -1 and -1.5 are plotted in the upper-right corner of each panel.}
    \label{SurfacaBrightness}
\end{figure*}

The post-perihelion images were used to characterize the large-scale radial surface-brightness distribution and its temporal evolution, because the greater extent of the outbound coma allowed the profiles to be traced over substantially larger distances. Each annulus had a projected width of \(10^3\) km, and the profiles were measured over projected distances from \(10^3\) to \(8\times10^4\) km. Azimuthally averaged radial surface-brightness profiles were extracted from the post-perihelion images using concentric annuli centered on the photometric center. Each annulus had a projected width of $10^{3}$~km, and the profiles were measured over projected cometocentric distances from $10^{3}$ to $8\times10^{4}$~km. The 1000~km annuli correspond to approximately 0.61--0.76 pixel over the post-perihelion observations and therefore oversample the matched PSF. We retain these annuli to sample and display the radial profiles, but exclude the region within $\rho < 1.5\,\mathrm{FWHM}_{\rm PSF}$ from the slope analysis. For the slope measurements, eight adjacent annuli were combined into 8000~km radial bins, corresponding to approximately 1.6--2.0 times the matched PSF FWHM. The sky background was estimated from 16 boxes, each with a size of $12\times12$ pixels, selected from regions that were not significantly affected by the coma or tail and were free from field-star contamination. Field stars and other contaminating sources were masked, and the remaining pixels in each sky box were sigma-clipped. The local sky background was taken as the mean of the median signals measured in these boxes, and the scatter among the boxes was included in the surface-brightness uncertainty. For each radial annulus, the total flux was measured after subtracting the sky contribution and was converted to surface brightness. Each profile of the surface brightness was then normalized to the innermost valid annulus and analyzed in \(\log \rho\) and \(\log S\) space, where \(\rho\) is the projected cometocentric distance and \(S(\rho)\) is the normalized surface brightness. The resulting profiles for the 16 post-perihelion epochs are shown in Fig.~\ref{SurfacaBrightness}, grouped into four panels of four epochs each to reduce crowding. 

In the standard idealized description of cometary comae, a spherically symmetric, steady dust outflow is often approximated by a logarithmic surface-brightness slope near \(-1\), while the outer profile can steepen toward \(-1.5\) as radiation pressure increasingly modifies the dust distribution \citep{jewitt1987surface}. The post-perihelion profiles show a broadly similar two-regime structure. The inner coma follows a relatively shallow profile close to the \(\rho^{-1}\) reference slope, while the outer coma steepens toward the \(\rho^{-1.5}\) reference slope. Notably, between 2025 November 25 and 2026 January 12, the outer profiles beyond \(\rho \sim 4\times10^4\) km become steeper than the \(\rho^{-1.5}\) reference slope. To quantify the outer-coma evolution, we fitted $S(\rho)\propto\rho^k$ over \(4\times10^4<\rho<8\times10^4~{\rm km}\). The azimuthally averaged slope, \(k_{\rm azi}\), changed from \(-2.78\) on UT 2025 November 25 to approximately \(-1.65\) to \(-1.78\) in January 2026. The fitted slopes are listed in Table~\ref{tab:haser_mixture_slopes}. After 2026 January 12, the outer profiles become shallower again and more closely follow the \(\rho^{-1.5}\) reference slope.

However, the slopes are derived from azimuthally averaged profiles,
and the asymmetric coma morphology shown in Fig. \ref{fig:ztf_sequence} may affect their interpretation. As the viewing geometry changes, an elongated dust structure can be
projected differently onto the sky plane. In particular, material at
large three-dimensional cometocentric distances may be shifted toward
smaller projected distances when the structure becomes more closely
aligned with the line of sight. This redistribution can modify the
azimuthally averaged radial profile and its fitted slope. To assess the sensitivity of the measured slopes to the asymmetric morphology, we additionally extracted profiles from two opposite sectors perpendicular to the projected Sun--comet axis and fitted their
outer slopes, \(k_{\perp}\), over the same radial interval. The corresponding perpendicular surface-brightness profiles are shown in Fig.~\ref{fig:transverse_profiles}. The
perpendicular profiles are generally steeper than the azimuthally
averaged profiles through UT 2025 December 20, indicating that the
asymmetric morphology affects the absolute value of the measured outer
slope. The excess steepening is therefore not an artifact introduced by the principal asymmetric structure, since it remains, and is even stronger, in the perpendicular profiles. Nevertheless, both profile sets exhibit temporal flattening. Specifically, \(k_{\perp}\) changes from \(-2.92\) on UT 2025 November 25 to approximately \(-1.32\) in January 2026, following the same overall temporal trend as \(k_{\rm azi}\)
(Table~\ref{tab:haser_mixture_slopes}).

We further tested whether the observed temporal evolution could be
produced solely by the changing viewing geometry. We adopted the
UT 2025 November 25 image as a fixed reference morphology and varied
only the phase-angle-dependent projection. The asymmetric component
was compressed along the projected Sun--comet axis according to $s_i=\frac{\sin\alpha_i}{\sin\alpha_0}$, where \(\alpha_0\) and \(\alpha_i\) are the phase angles of the reference and simulated epochs, respectively. Each simulated image was
then azimuthally averaged, and \(k_{\rm proj}\) was fitted over the same
radial interval as the observations. In contrast to the observed
flattening, the model predicts progressive steepening as the phase
angle decreases, reaching \(k_{\rm proj}\sim-3.25\) in January
(Table~\ref{tab:haser_mixture_slopes}). This behavior is consistent with a simple physical picture in which an elongated antisolar dust structure becomes increasingly aligned with the line of sight as the phase angle decreases, producing a steeper projected radial profile. In contrast, both the observed azimuthally averaged and perpendicular profiles become substantially shallower. Thus, within this simple geometry, changes in viewing geometry alone do not explain the observed temporal flattening.

The persistence of the excess steepening in the perpendicular profiles,
together with the failure of the fixed-morphology projection model to
reproduce the observed evolution, suggests that the behavior reflects
physical evolution within the dust coma. Possible causes include the
fading or destruction of grains through sublimation or fragmentation,
or temporal variations in dust production that deplete the outer coma
relative to a steady outflow \citep{jewitt1987surface}.

\section{Discussion} \label{Dicussion}
\subsection{Heliocentric index} \label{Heliocentric index}
\begin{figure*}
\centering

\begin{minipage}{0.49\textwidth}
    \centering
    \includegraphics[width=\textwidth]{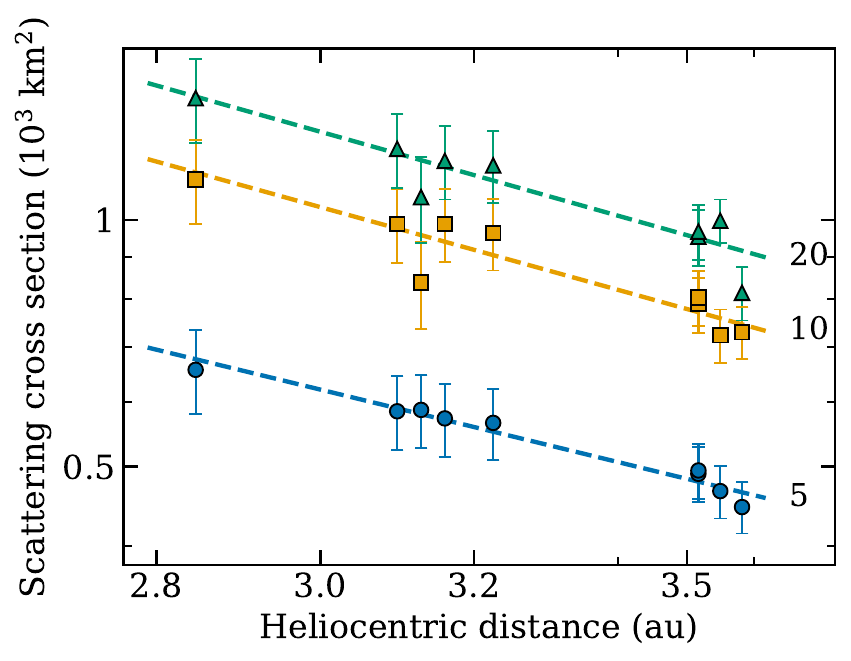}   
\end{minipage}
\hfill
\begin{minipage}{0.49\textwidth}
    \centering
    \includegraphics[width=\textwidth]{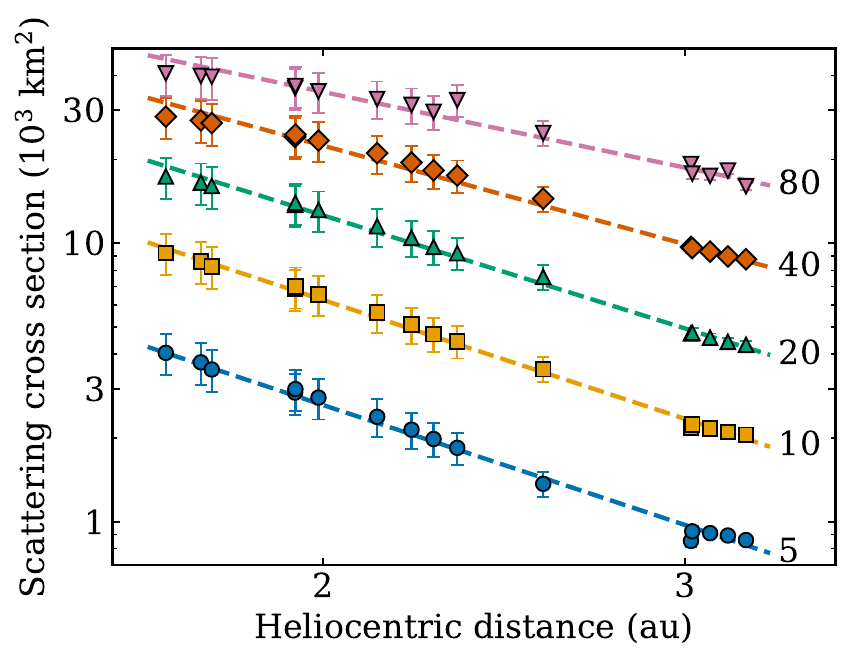} 
\end{minipage}

\caption{Effective scattering cross section as a function of heliocentric distance measured in projected photometric apertures of different radii. The left panel shows the pre-perihelion data, and the right panel shows the post-perihelion data. Symbols show the measured values, and dashed lines show power-law fits of the form \(C_{\rm e}=C_0 r_{\rm H}^{-\gamma}\). The numbers next to the fitted curves indicate the projected aperture radii in units of \(10^3\) km.}
\label{crossse}
\end{figure*}

We characterize the heliocentric behavior of the coma brightness by fitting the effective scattering cross section with a power law of $C_{\rm e}(r_H)=C_0 r_H^{-\gamma}$, where \(C_0\) is a normalization constant and \(\gamma\) is the heliocentric index. The fits were performed separately for the pre- and post-perihelion data in each projected photometric aperture. The results are shown in Fig.~\ref{crossse}. Before perihelion, the fitted index increases with aperture size. We obtain $\gamma=1.63\pm0.14$, $1.86\pm0.27$, and $1.88\pm0.34$ for projected aperture radii of $5\times10^3$, $1\times10^4$, and $2\times10^4$ km, respectively. Thus, the effective scattering cross section in the outer part of the sampled pre-perihelion coma exhibits a steeper dependence on heliocentric distance than that measured closer to the nucleus. The post-perihelion behavior shows the opposite aperture dependence. The fitted heliocentric index decreases systematically with increasing aperture size, from $\gamma=2.44\pm0.09$ at $5\times10^3$ km to $2.42\pm0.06$, $2.30\pm0.04$, $2.01\pm0.04$, and $1.54\pm0.09$ at $1\times10^4$, $2\times10^4$, $4\times10^4$, and $8\times10^4$ km, respectively. Thus, the outer coma exhibits a steeper inbound rise but a shallower outbound decline than the inner coma.

The contrast between the pre- and post-perihelion measurements indicates that the coma is not well described by a single aperture-independent heliocentric law across the full observing interval. Before perihelion, $\gamma$ increases with aperture within the inner $\sim2\times10^4$ km, indicating a stronger apparent heliocentric dependence in the outer part of the sampled coma. After perihelion, by contrast, $\gamma$ decreases systematically with increasing aperture, so that the outer coma exhibits a weaker apparent dependence on heliocentric distance than the region close to the nucleus. This aperture dependence suggests a differential temporal response between the inner and outer coma in the post-perihelion phase. 

\begin{table*}
    \centering
    \caption{Heliocentric indices from photometry of comet 3I}
    \label{tab:heliocentric_indices}
    
    \renewcommand{\arraystretch}{1.25} 
    
    \setlength{\tabcolsep}{15pt} 

    \begin{tabular}{lcccc}
        \toprule
        Phase$^a$ & $r_H$$^b$ (au) & Photometry$^c$ & $n$$^d$ & References \\
        \hline
        Pre  & 3.6--2.8 & fixed radius, $5\times10^3$ km & $3.6 \pm 0.1$ & [1] \\
        Pre  & 4.4--2.1 & fixed radius, $7\times10^4$ km & $5.1 \pm 0.1$ & [2] \\
        Pre  & 4.6--1.8 & fixed radius, $10^4$ km        & $3.8 \pm 0.3$ & [3] \\
        Pre  & 7.7--2.6 & fixed angle, 10\arcsec         & $4.2 \pm 0.4$ & [4] \\
        Pre  & 6.5--3.7 & fixed angle, 3\arcsec          & $3.9 \pm 0.1$ & [5] \\
        Pre  & 6.3--5.5 & fixed angle, 63\arcsec         & $10.4 \pm 1.2$& [6] \\
        Post & 1.7--3.2 & fixed radius, $5\times10^3$ km & $4.4 \pm 0.1$ & [1] \\
        Post & 1.8--3.3 & fixed radius, $10^3$ km        & $4.5 \pm 0.3$ & [7] \\
        \hline
    \end{tabular}
    
    \vspace{1.0em} 
    
    \begin{minipage}{0.7\textwidth} 
        \footnotesize
        \linespread{1.3}\selectfont 
        
        \textit{Notes.} \\
        $^a$ Orbital phase: pre-perihelion and post-perihelion. \\
        $^b$ Heliocentric distance range of the photometry, in au. \\
        $^c$ Photometric measurement type and aperture size. \\
        $^d$ Magnitude-based heliocentric index. \\
        \textbf{References.} [1] This work; [2] \cite{hartman2026hatpi}; [3] \cite{jewitt2025preperihelion}; [4] \cite{tonry2025atlas}; [5] \cite{ye2025prediscovery}; [6] \cite{martinez2025prediscovery}; [7] \cite{hui2026nucleus}.
    \end{minipage}
\end{table*}

A possible interpretation is that different apertures sample dust released over different characteristic timescales. Small apertures are weighted more strongly toward recently released material and may therefore track the current activity level more directly. Larger apertures also include dust emitted at earlier epochs and retained over longer residence times, so that the measured cross section may reflect a more time-averaged activity state. Because this time-averaging is always present, the effect becomes observable only when the underlying activity varies rapidly enough for dust of different ages to carry appreciably different signals. This picture may help explain the difference between the pre- and post-perihelion branches of the $C_{\rm e}$--$r_{\rm H}$ relation. In this sense, the post-perihelion aperture dependence is consistent with a scale-dependent lag effect, whereas the trend before perihelion may additionally reflect the progressive filling of the outer coma, grain fragmentation, or an extended grain source whose relative contribution increases with cometocentric distance.

This interpretation may also help explain the asymmetry between the pre- and post-perihelion slopes measured in the largest aperture. If the cross section in the outermost aperture is treated as a proxy for the integrated dust reservoir of the coma, then a shallower post-perihelion variation can arise naturally from temporal averaging. Before perihelion, the large-aperture flux averages over material released at earlier and generally fainter stages of activity, whereas after perihelion it still contains material emitted near the epoch of peak activity and therefore retains a stronger memory of a previously brighter state. The effect could be further enhanced if icy grains contribute to the cross section on large scales. In this respect, our result is qualitatively consistent with the large-aperture gas asymmetry reported by \citet{tan2026perihelion}, who found a steeper inbound rise than outbound decline in \(Q({\rm H_2O})\) and interpreted it as the signature of an extended icy-grain source that built up toward perihelion and did not undergo rapid post-perihelion depletion.

At the opposite limit, the smallest aperture is least affected by long-residence-time material and therefore most nearly traces the contemporaneous state of the inner coma. The difference between the pre- and post-perihelion indices measured in the \(5\times10^{3}\) km aperture likely reflects a real change in the physical conditions of the near-nucleus dust environment. In this aperture, we find $\gamma_{\rm pre}=1.63\pm0.14$ and \(\gamma_{\rm post}=2.44\pm0.09\), corresponding to magnitude-based heliocentric indices of \(n_{\rm pre}=3.63\pm0.14\) and \(n_{\rm post}=4.44\pm0.09\) through the relation \(n=\gamma+2\). To place these values in context, Tab.~\ref{tab:heliocentric_indices} summarizes published magnitude-based heliocentric indices for comet 3I derived from photometry. Our pre-perihelion index is consistent with the result of \citet{jewitt2025preperihelion} and, within the uncertainties, also with those reported by \citet{tonry2025atlas,ye2025prediscovery}. By contrast, the steeper pre-perihelion indices reported by \citet{hartman2026hatpi,martinez2025prediscovery} may be related to the use of larger photometric apertures. \citet{hartman2026hatpi} suggested that larger apertures include a more extended component whose composition and scattering behavior may differ from those of the inner coma and which may also contain a larger gas contribution. If that interpretation is applicable here, the larger apertures would sample material whose response to solar heating and radiation is stronger than that of the near-nucleus dust, thereby steepening the measured heliocentric index. Our post-perihelion index is consistent with the Hubble result reported by \citet{hui2026nucleus}, which supports the overall reliability of our measurements in both orbital phases.

The pre-perihelion index measured in our smallest aperture is also consistent with the result of \citet{jewitt2025preperihelion}, who showed that the observed heliocentric dependence is compatible with sublimation driven by CO or CO\(_2\), but is shallower than expected for H\(_2\)O over the same heliocentric range. Given the independent spectroscopic identification of CO$_2$ as the dominant volatile at \(r_H \sim 3.3\) au, the simplest interpretation is that the onset of detectable activity was driven primarily by CO$_2$ \citep{jewitt2025preperihelion}. The heliocentric index alone does not uniquely distinguish CO$_2$ from CO, and some contribution from CO cannot be excluded. The steeper post-perihelion index suggests that the activity evolved onto a different volatile regime on the outbound leg. \citet{hui2026nucleus} noted that 3I was CO$_2$-dominated at \(r_H \gtrsim 3\) au and became H$_2$O-dominated closer to the Sun. In that context, our post-perihelion result is consistent with a stronger contribution from H\(_2\)O after perihelion. However, we caution that the heliocentric index is not a unique diagnostic, and that some part of the steepening may also reflect evolution in the dust environment.

\subsection{Perpendicular Velocity}
\label{subsec:vperp}
The viewing geometry on 2026 January 18 provides a direct constraint on the out-of-plane motion of the dust. At this epoch, the line of sight was only $\sim0.88^\circ$ from the orbital plane of 3I, so projection effects on the apparent vertical thickness of the dust sheet were small. The transverse tail width, characterized by its full width at half maximum (FWHM), can therefore be used as an approximate measure of the dust distribution perpendicular to the orbital plane. Fig.~\ref{FWHM_two_sides_signed} shows the relationship between the tail FWHM and the distance from the nucleus. We measured the observed tail width, $w_{\rm obs}$, defined as the FWHM of surface-brightness profiles extracted perpendicular to the projected tail axis. Each profile was averaged over a finite segment along the tail axis to improve the signal-to-noise ratio.

Following \cite{jewitt2014hubble}, we interpret the width--distance relation in the tail as arising from simultaneous dust motion along the tail and perpendicular to the orbital plane. A dust grain moving with speed \(v_\perp\) perpendicular to the orbital plane and accelerated along the tail by solar radiation pressure travels a projected distance of $\ell_T=(1/2)\beta g_\odot t^2$, where $\beta$ is the ratio of radiation-pressure acceleration to solar gravity and $g_\odot=g_1/r_H^2$ is the solar gravitational acceleration at heliocentric distance $r_H$ \citep{liu2025great}. Taking the measured width to trace the separation between particles moving to either side of the orbital plane and eliminating $t$ gives $w_T = v_\perp\left(8\ell_T/{\beta g_\odot}\right)^{1/2}$,
where $w_T$ is the dynamical contribution to the observed tail width. Because the image is seeing-limited, the measured FWHM includes both the dynamical broadening and an unresolved width near the nucleus. We therefore write $w_{\rm obs}^2 = w_0^2 + w_T^2$,
where $w_0$ is an effective width that accounts for seeing and unresolved near-nucleus structure. With radiation-pressure-driven motion along the tail, this yields
\begin{equation}
w_{\rm obs}=\left[w_0^2+\left(v_{\perp}\sqrt{\frac{8\ell_T}{\beta g_\odot}}\right)^2\right]^{1/2}.
\label{eq:tail_width_final}
\end{equation}
The measurements on the tail and anti-tail sides in Fig.~\ref{FWHM_two_sides_signed} were fitted separately with Eq.~\ref{eq:tail_width_final} to allow for possible asymmetry. The fits give perpendicular velocities of $v_{\perp} =(73\pm5)\,\beta^{1/2}\ {\rm m\ s^{-1}}$ for the tail side and $v_\perp =(85\pm7)\,\beta^{1/2}\ {\rm m\ s^{-1}}$ for the anti-tail side. Given the similarity of the two fitted values, we adopt a representative perpendicular velocity of $v_\perp = (79\pm6)\,\beta^{1/2}\ {\rm m\ s^{-1}}$. Following \cite{liu2024unraveling}, we adopt the approximation $\beta \approx a_{\mu{\rm m}}^{-1}$, where $a_{\mu{\rm m}}$ is the grain radius in microns. The representative perpendicular velocity then becomes $v_\perp \approx 79\,a_{\mu{\rm m}}^{-1/2}\ {\rm m\ s^{-1}}$. For example, a grain with $a=100~\mu{\rm m}$ has $\beta \approx 0.01$ and hence $v_{\perp,0} \approx 7.9\ {\rm m\ s^{-1}}$. 


The representative perpendicular velocity derived here is about an order of magnitude larger than the pre-perihelion HST value, \(v_\perp=7.5\,\beta^{1/2}~{\rm m~s^{-1}}\), obtained at \(r_{\rm H}=3.83\) au by \citet{jewitt2025hubble}. Because \(v_\perp\) is inferred from the transverse tail width, this difference is better interpreted as a change in the effective out-of-plane dust dispersion, possibly driven by a change in source geometry. Before perihelion, \citet{serra2025pre} interpreted the morphology as a single high-latitude jet precessing around the rotation axis, whereas post-perihelion HST images show multiple faint, straight jet-like structures and a more fan-like coma \citep{manzini2025jets}. Together with spectroscopic evidence that different gas species may trace distinct coma structures \citep{hoogendam2025university}, this suggests that multiple active regions or newly illuminated jets at different position angles could broaden the ejecta distribution and increase the effective \(v_\perp\).

Combining the sunward ejection speed inferred from the turnaround distance in Sect.~\ref{Morphology} with the perpendicular velocity, we estimate the characteristic opening angle of the post-perihelion ejection cone following $\theta \sim 2\tan^{-1}\left({v_\perp}/{v_{\rm ej}}\right)$. Using $v_\perp=(79\pm6)\beta^{1/2}\ {\rm m\ s^{-1}}$ and
$v_{\rm ej}=(190\pm3)\beta^{1/2}\ {\rm m\ s^{-1}}$, we obtain $\theta \sim 45\pm3^\circ$. This value is substantially larger than the pre-perihelion fan opening angle of $\sim17^\circ$ inferred by \citet{jewitt2025hubble}, suggesting that the post-perihelion dust had a broader distribution of emission directions than at the pre-perihelion epoch.

\subsection{Mass loss rate}
\label{Mass loss rate}
The observed tail length provides a constraint on the minimum effective particle size. We measured this length from the image obtained on 28 November 2025. The tail axis was determined from the peak of the smoothed azimuthal median profile, giving a position angle of \(164^\circ\) in image coordinates. An approximately axisymmetric coma model was constructed after excluding the region near the tail axis, and subtracted from the original image to form a residual image. A one-dimensional surface-brightness profile was then extracted along the tail axis. We defined the tail length as the outermost distance at which two consecutive bins satisfied \(\mathrm{S/N}>4\), and obtained a projected length of \(210\arcsec\), equal to $\sim 3\times10^5$ km. 

According to \citet{ye2025prediscovery}, sustained dust outflow from comet 3I probably began in early April 2025, at a heliocentric distance of \(\sim 9\) au. Adopting this onset time, the dust would have been released \(\sim2.1\times10^7\) s before the observation, and its radiation-pressure displacement is
\begin{equation}
L \sim \frac{\beta g_{\odot}(1)}{2r_{H,\mathrm{ej}}^{2}}\,t^{2}.
\end{equation}
Here, \(g_{\odot}(1)=0.006~\mathrm{m~s^{-2}}\) is the solar gravity at 1 au, \(r_{H,\mathrm{ej}}\) is the heliocentric distance at ejection, and \(\beta\) is the radiation-pressure parameter. With \(L\sim 3\times10^{5}\) km in projection, we infer an upper limit to \(\beta\), and hence a lower limit to the effective particle size. Using \(\beta \sim a_{\mu\mathrm{m}}^{-1}\), we find $a_{\min} \sim 55\,\mu\mathrm{m}$.
Owing to projection effects and the time dependence of \(r_H^{-2}\), the derived size should be regarded as an order-of-magnitude lower limit. 

The upper limit of the dust size is less directly constrained by the image morphology. We therefore estimate a maximum ejectable radius by equating the gas drag force to the weight of a spherical particle. Following \citet{jewitt2019distant}, this gives $a_{\rm max}={9 C_D V_g f_s(r_H)}/
{16\pi G \rho_d \rho_n r_n}$,
where the drag coefficient is set to \(C_D=1\), the gas speed is adopted from \cite{cordiner2026isotopic} as \(V_g=310~\mathrm{m~s^{-1}}\), the dust and nucleus densities are set to \(\rho_d=\rho_n=500~\mathrm{kg~m^{-3}}\), and the nucleus radius is adopted from \cite{hui2026nucleus} as \(r_n=1.3~\mathrm{km}\). In the heliocentric distance range of our observations, \(1.6 \lesssim r_H \lesssim 3.6\) au, we estimate the sublimation mass flux using the thermal energy balance equation,
\begin{equation}
\frac{F_{\odot}}{r_H^{2}}(1-A)=\chi\left[\epsilon \sigma T\left(r_H\right)^{4}+f_{s} L_{s}\right],
        \label{waterRate}
\end{equation}
where $F_{\odot}=1361$ W m$^{-2}$ is the solar constant, \(r_H\) is the heliocentric distance in au, $A=0.04$ is the assumed Bond albedo, $\epsilon=0.9$ is the assumed infrared emissivity, $\sigma = 5.67\times10^{-8}$ W m$^{-2}$ K$^{-4}$ is the Stefan--Boltzmann constant, $\chi=2$ is the adopted heat-redistribution factor corresponding to reradiation and sublimation of the absorbed solar energy over the sunlit hemisphere, and $L_s=2.8\times10^6~\mathrm{J~kg^{-1}}$ is the latent heat for water ice. Substitution into Eq.~\ref{waterRate} gives \(f_s \sim 1.8\times10^{-6}\text{--}6.5\times10^{-5}~\mathrm{kg~m^{-2}~s^{-1}}\), with the larger value at smaller \(r_H\). Over \(1.6\lesssim r_H\lesssim3.6\) au, this corresponds to \(a_{\rm max}\sim5\)--\(166\) mm. We adopt the lower end of this range as a conservative representative value, \(a_{\rm max}\sim5~{\rm mm}\). This value represents the maximum particle radius that can be lifted across the full observing interval. Larger particles could still be ejected at smaller \(r_H\), where the sublimation flux is higher.

In the absence of a direct constraint on the dust-size distribution, we adopt the commonly used differential size distribution \(n(a)\,da\propto a^q\,da\) with \(q=-3.5\), corresponding to the canonical collisional-equilibrium value \citep{dohnanyi1969collisional,o2003steady} and also adopted in cometary dust analyses of 2I/Borisov \citep{kim2020coma}. For this distribution, the cross-section-weighted mean particle radius is $\bar{a}=(a_0a_1)^{1/2}$. Taking \(a_{\rm min}\sim55~\mu{\rm m}\) and \(a_{\rm max}\sim5~{\rm mm}\), we obtain \(\bar{a}\sim0.5~{\rm mm}\), corresponding to \(\beta\sim2\times10^{-3}\) for \(\beta= a_{\mu{\rm m}}^{-1}\). This value lies within the \(\beta\approx10^{-3}\)--\(10^{-2}\) range inferred from the Tianwen-1 syndyne--synchrone modeling by \citet{ren2026interstellar}, and is broadly consistent with the large-grain interpretation of comet 3I. It is also of the same order of magnitude as the \(\sim100~\mu{\rm m}\) effective size adopted in pre-perihelion analyses of \cite{jewitt2025preperihelion}, and remains within the hundred-micron to millimeter regime suggested by \citet{bolin2025interstellar,scarmato2025interstellar}.

We then estimate the dust production rate from the effective scattering cross section. Following \citet{liu2024unraveling}, the dust mass within an aperture is $M_d=\frac{4}{3}\rho_d\bar{a}C_e$, where \(C_e\) is the effective scattering cross section. The corresponding residence-time averaged dust production rate is $\dot{M}_d={M_d}/{\tau_r}$, where \(\tau_r=R_{\rm ap}/v_{\rm ej}\) is the residence time of dust grains in an aperture of projected radius \(R_{\rm ap}\). We adopt the largest photometric aperture at each epoch in order to include as much of the dust coma as possible. The aperture radius is \(R_{\rm ap}=20000\) km for the pre-perihelion measurements and \(R_{\rm ap}=80000\) km for the post-perihelion measurements. Although the ejection-speed relation was derived from the pre-perihelion sunward-turnaround measurements, the same velocity scaling is applied to both branches for consistency. The resulting dust mass-loss rates should therefore be interpreted as order-of-magnitude, residence-time averaged estimates. Using \(\bar{a}\sim0.5~{\rm mm}\), \(\rho_d=500~{\rm kg~m^{-3}}\), and the ejection-speed relation of \(v_{\rm ej}=(190\pm3)\beta^{1/2}~{\rm m~s^{-1}}\) derived in Sect.~\ref{Morphology}, the adopted mean radius corresponds to \(\beta\sim2\times10^{-3}\) and hence \(v_{\rm ej}\sim8.5\pm0.4~{\rm m~s^{-1}}\). We adopt \(v_{\rm ej}\sim9~{\rm m~s^{-1}}\) for the residence-time estimate, giving \(\tau_r\sim2.2\times10^6\) s for the pre-perihelion aperture and \(\tau_r\sim8.9\times10^6\) s for the post-perihelion aperture. 

Using the effective scattering cross sections listed in Tab.~\ref{tab:cross}, we computed the residence-time averaged dust production rates of \(\dot{M}_d\sim123\)--\(203~{\rm kg~s^{-1}}\) before perihelion and \(\dot{M}_d\sim596\)--\(1412~{\rm kg~s^{-1}}\) after perihelion. The resulting values are plotted as a function of heliocentric distance in Fig.~\ref{dustloss}. Fits to the pre- and post-perihelion measurements give \(\dot{M}_d\propto r_H^{-1.88\pm0.34}\) and \(\dot{M}_d\propto r_H^{-1.54\pm0.09}\), respectively. Since \(\dot{M}_d\propto C_e\) for the adopted parameters, these indices are identical to those measured from the largest-aperture scattering cross sections in Sect.~\ref{Heliocentric index}. Extrapolating these fitted relations to the perihelion distance \(q=1.35\) au gives \(\dot{M}_{d,{\rm pre}}=849\pm244~{\rm kg~s^{-1}}\) and \(\dot{M}_{d,{\rm post}}=2164\pm84~{\rm kg~s^{-1}}\). These extrapolated values should be interpreted with caution. The pre-perihelion estimate is less secure because the last pre-perihelion measurement is farther from \(q\) than the first post-perihelion measurement, requiring a longer extrapolation. Moreover, the activity may deviate from a single power law near perihelion, where small changes in \(r_H\) can strongly affect the sublimation-driven dust production.


We further use the fitted mass-loss laws to estimate the cumulative dust mass released between the largest observed heliocentric distance on each branch and perihelion. For each branch, we integrated \(\dot{M}_d(t)=\dot{M}_0 r_H(t)^k\) over time using the daily heliocentric distances from JPL/Horizons\footnote{\url{https://ssd.jpl.nasa.gov/horizons/}} and the corresponding pre- or post-perihelion fit. This gives \(M_{\rm d,pre}\sim3.3\times10^9~{\rm kg}\) for the inbound branch and \(M_{\rm d,post}\sim8.8\times10^9~{\rm kg}\) for the outbound branch, for a total of \(M_{\rm d,tot}\sim1.2\times10^{10}~{\rm kg}\). Expressed as an equivalent sphere with density \(\rho_d=500~{\rm kg~m^{-3}}\), the radius corresponding to a cumulative dust mass \(M_d\) is $R_{\rm eq}=\left({3M_d}/{4\pi\rho_d}\right)^{1/3}$. Substitution gives \(R_{\rm eq}\sim180\) m for the total dust loss. If contamination from gas emission or other non-dust sources within the photometric aperture is negligible, this equivalent radius provides a mass-budget lower limit to the pre-activity nucleus radius.

\subsection{Deviation of surface brightness}
\label{Deviation of surface brightness}

In a simple steady dust coma, the surface brightness is expected to scale approximately as $S(\rho)\propto \rho^{-1}$ in the inner coma, while radiation pressure can steepen the outer profile toward $\rho^{-3/2}$ \citep{jewitt1987surface}. For comet 3I, the post-perihelion surface-brightness profiles in Fig.~\ref{SurfacaBrightness} show a transient excess steepening relative to this outer-coma reference. We measured the effect by fitting a single logarithmic slope over $4\times10^4 < \rho < 8\times10^4~{\rm km}$. The resulting values of $k$ are given in Tab.~\ref{tab:haser_mixture_slopes}. The deviation was strongest in the earliest post-perihelion epochs. On UT 2025 November 25, we find $k=-2.78$, and similarly steep values persist through early December. The profile then  flattened with time, reaching $k=-1.95$ on UT 2025 December 28 and $k\sim -1.65$ to $-1.78$ during UT 2026 January 12--18. 

One might suspect that the steep outer profile reflects a recent increase in dust production. If the comet were ejecting more dust at the time of observation than at earlier times, the younger dust in the inner coma would be brighter than the older dust farther out, thereby steepening the radial profile. This effect is possible in principle, but it is not favored by our photometry. During the post-perihelion phase, $C_{\rm e}$ in Tab.~\ref{tab:cross} decreases in all apertures, with the smallest apertures fading most rapidly. Thus the steep outer profile occurs while the near-nucleus scattering material is being depleted, not replenished. A recent increase in dust production is therefore unlikely to be the origin of the transient steepening. 

Fragmentation is another process capable of modifying the surface-brightness profile, but its expected photometric signature is generally opposite to the excess steepening observed in comet 3I. For approximately fixed dust mass, fragmentation into daughter grains that remain optically efficient scatterers increases the total optical cross-section, since cross-section scales as $a^2$ whereas mass scales as $a^3$. Such fragmentation would brighten the outer coma relative to the inner coma and would therefore tend to flatten the profile. Fragmentation all the way down to particles much smaller than the observing wavelength could in principle steepen the profile, because the scattering efficiency drops rapidly in the Rayleigh regime. However, by itself this mechanism does not explain the recovery of the outer slope toward $\rho^{-3/2}$ by the January epochs. This would require the fragmentation efficiency to decline rapidly with increasing heliocentric distance, implying an additional temperature-dependent process.

\begin{figure}
\centering
    \includegraphics[width=0.6\columnwidth]{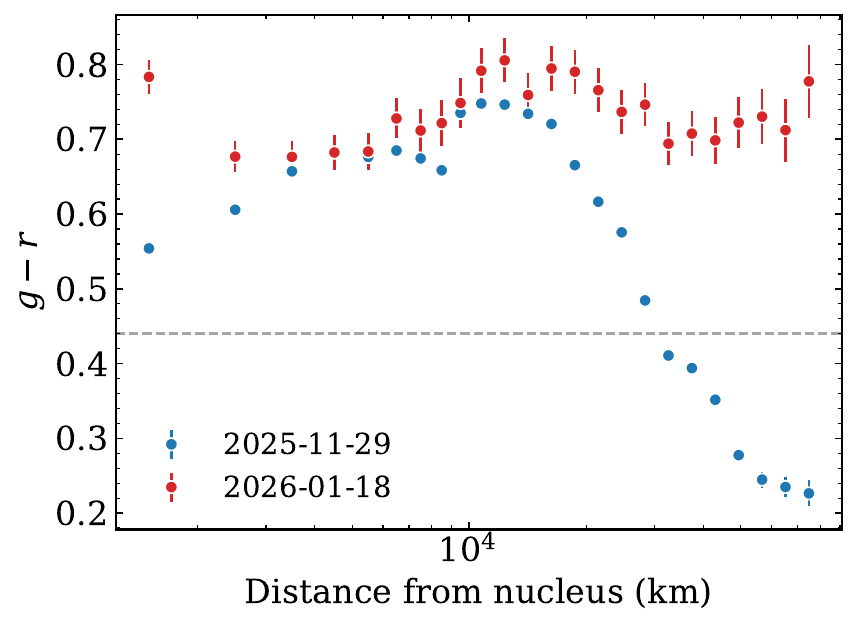}
    \caption{Radial \(g-r\) color profiles of 3I on 2025 November 29 and 2026 January 18. The horizontal dashed line marks the solar \(g-r\) color.}
    \label{color}
\end{figure}

The observed behavior is more naturally described as grain fading. In this scenario, grains contribute to the inner-coma scattering but lose optical cross-section as they move outward. Sublimation of icy grains provides a plausible mechanism, because their loss rates depend sensitively on heliocentric distance. The resulting decline in scattering efficiency steepens the surface-brightness profile, with freshly released grains dominating near the nucleus and faded grains contributing farther out. This mechanism can account for the outer-coma steepening seen in the early post-perihelion epochs. The later flattening by January is consistent with a weakening fading component as the supply of sublimating icy grains decreases at larger heliocentric distances. This interpretation is supported more broadly by independent reports of icy material and distributed icy-grain sources in the coma of 3I \citep{yang2025spectroscopic,lisse2026spherex}.

The color profiles provide an independent check on the fading interpretation. We compared radial \(g-r\) profiles measured on UT 2025 November 29, when the outer \(r\)-band profile was still relatively steep, with those measured on UT 2026 January 18, after the outer profile had flattened toward the nominal \(\rho^{-3/2}\) behavior. For each epoch, the color profile was derived from paired ZTF \(g\)- and \(r\)-band images obtained within 2 hr of each other, which reduces the influence of temporal evolution between the two filters. The two images were analyzed using the same photometric center, projected annuli, and sky-background estimation as adopted for the surface-brightness profiles. To minimize seeing-induced color gradients, the image with the narrower point-spread function was convolved with a Gaussian kernel to match the broader point-spread function before extracting the radial profiles. The \(g-r\) profile was then computed from the difference between the calibrated \(g\)- and \(r\)-band surface-brightness profiles, and the uncertainty was obtained by propagating the independent errors in the two bands. The resulting profiles are shown in Fig.~\ref{color}. On UT 2025 November 29, the color increases from $g-r=0.63\pm0.01$ in the $1\times10^3$--$5\times10^3~{\rm km}$ annulus to $g-r=0.74\pm0.01$ in the $1\times10^4$--$2\times10^4~{\rm km}$ annulus. At larger radii, where PSF effects are less important, the profile turns sharply blueward, reaching $g-r=0.30\pm0.02$ over $4\times10^4$--$8\times10^4~{\rm km}$. By UT 2026 January 18, the strong radial color structure has largely disappeared. The profile is redder and nearly flat, with $g-r=0.71\pm0.02$ in the $1\times10^3$--$5\times10^3~{\rm km}$ annulus and $g-r=0.72\pm0.05$ over $4\times10^4$--$8\times10^4~{\rm km}$. 

The disappearance of the color structure coincides with the recovery of the outer surface-brightness profile. One possible interpretation is that the early post-perihelion coma contained a short-lived scattering component. In the November profile, its signature is most evident at large projected radii, where the $g-r$ color becomes anomalously blue over roughly the same radial range in which the $r$-band surface-brightness profile is unusually steep. If these grains lose optical cross-section as they move outward, the outer coma is depleted relative to a non-fading dust flow, producing a profile steeper than the nominal $\rho^{-3/2}$ behavior. In the January profile, this component is no longer prominent within the measured radial range. The grains responsible for the November anomaly may have lost their scattering cross-section, moved beyond the measured radial range, or been supplied at a lower rate as the comet receded from the Sun. The color profile then becomes redder and flatter, while the surface-brightness slope relaxes toward reference dust-coma behavior.

To measure the fading scale, we fitted the early post-perihelion profiles with an empirical projected Haser-type destruction component, motivated by \citet{keto2026physics}. We treated the average January profile as the late-time baseline and fitted each earlier normalized profile with
\begin{equation}
P_{\rm model}(\rho)
=
\frac{
P_{\rm base}(\rho)
+\eta\, H_{\rm norm}(\rho;\ell_{\rm DE})
}
{1+\eta}.
\label{P_model}
\end{equation}
Here \(P_{\rm base}\) is the normalized January baseline, \(\eta\) is the
relative weight of the fading component, and \(\ell_{\rm DE}\) is the
projected destruction length. We used the projected Haser-type destruction
kernel of $H(\rho;\ell_{\rm DE})={K_0(\rho/\ell_{\rm DE})}/{\rho}$, following the surface-brightness form adopted by \citet{keto2026physics},
where \(K_0\) is the modified Bessel function of the second kind. In the
fit, this kernel was normalized at the innermost common radius of the
normalized profiles, so that \(H_{\rm norm}(\rho;\ell_{\rm DE})=H(\rho;\ell_{\rm DE})/H(\rho_{\rm norm};\ell_{\rm DE})\). The normalization radius \(\rho_{\rm norm}\) is the innermost common radius of the normalized profiles. For each earlier epoch, the fit yields the fading-component weight \(\eta\) and the effective projected destruction length \(\ell_{\rm DE}\). The fitted parameters are reported in Tab.~\ref{tab:haser_mixture_slopes}.

\begin{table*}
\centering
\renewcommand{\arraystretch}{1.15}

\caption{Observed and modeled outer surface-brightness slopes and
Haser-mixture fading lengths.}
\label{tab:haser_mixture_slopes}
\setlength{\tabcolsep}{15pt}
\begin{tabular}{lccccc}
\hline
UT Date &
\(k_{\rm azi}\) &
\(k_{\perp}\) &
\(k_{\rm proj}\) &
\(\eta\) &
\(\ell_{\rm DE}\) \((10^4~{\rm km})\) \\
\hline
2025 Nov 25 12:07 &
\(-2.78\pm0.13\) & \(-2.92\pm0.21\) &
\(-2.78^{+0.00}_{-0.00}\) &
\(1.86^{+0.07}_{-0.05}\) &
\(6.45^{+0.76}_{-0.55}\) \\

2025 Nov 28 13:15 &
\(-2.71\pm0.19\) & \(-2.94\pm0.16\) &
\(-2.78^{+0.01}_{-0.00}\) &
\(1.64^{+0.06}_{-0.04}\) &
\(6.54^{+0.92}_{-0.60}\) \\

2025 Nov 29 11:38 &
\(-2.65\pm0.14\) & \(-2.90\pm0.23\) &
\(-2.78^{+0.01}_{-0.00}\) &
\(0.91^{+0.02}_{-0.02}\) &
\(6.49^{+0.87}_{-0.56}\) \\

2025 Dec 06 12:40 &
\(-2.72\pm0.12\) & \(-2.95\pm0.21\) &
\(-2.78^{+0.00}_{-0.00}\) &
\(0.92^{+0.02}_{-0.02}\) &
\(6.70^{+0.87}_{-0.59}\) \\

2025 Dec 06 13:12 &
\(-2.72\pm0.16\) & \(-2.96\pm0.23\) &
\(-2.78^{+0.00}_{-0.00}\) &
\(1.23^{+0.04}_{-0.03}\) &
\(6.77^{+0.76}_{-0.54}\) \\

2025 Dec 08 11:55 &
\(-2.74\pm0.15\) & \(-2.93\pm0.19\) &
\(-2.77^{+0.00}_{-0.01}\) &
\(0.97^{+0.03}_{-0.03}\) &
\(6.74^{+0.82}_{-0.57}\) \\

2025 Dec 13 11:57 &
\(-2.55\pm0.13\) & \(-2.92\pm0.25\) &
\(-2.79^{+0.00}_{-0.04}\) &
\(0.45^{+0.02}_{-0.02}\) &
\(6.79^{+0.80}_{-0.57}\) \\

2025 Dec 16 11:32 &
\(-2.48\pm0.13\) & \(-2.92\pm0.24\) &
\(-2.82^{+0.00}_{-0.07}\) &
\(0.48^{+0.02}_{-0.02}\) &
\(6.85^{+0.78}_{-0.50}\) \\

2025 Dec 18 09:56 &
\(-2.41\pm0.26\) & \(-2.75\pm0.33\) &
\(-2.86^{+0.00}_{-0.10}\) &
\(0.34^{+0.02}_{-0.02}\) &
\(7.02^{+0.80}_{-0.61}\) \\

2025 Dec 20 12:33 &
\(-2.31\pm0.20\) & \(-2.83\pm0.37\) &
\(-2.91^{+0.00}_{-0.13}\) &
\(0.42^{+0.02}_{-0.02}\) &
\(7.19^{+0.85}_{-0.58}\) \\

2025 Dec 28 10:23 &
\(-1.95\pm0.29\) & \(-2.47\pm0.30\) &
\(-3.18^{+0.03}_{-0.22}\) &
\(0.17^{+0.03}_{-0.02}\) &
\(17.88^{+13.34}_{-6.68}\) \(^{\dagger}\) \\

2026 Jan 12 06:47 &
\(-1.59\pm0.24\) & \(-1.51\pm0.35\) &
\(-3.26^{+0.09}_{-0.32}\) &
-- & -- \\

2026 Jan 12 10:06 &
\(-1.75\pm0.31\) & \(-1.66\pm0.37\) &
\(-3.26^{+0.09}_{-0.32}\) &
-- & -- \\

2026 Jan 14 08:15 &
\(-1.82\pm0.33\) & \(-1.55\pm0.42\) &
\(-3.25^{+0.09}_{-0.34}\) &
-- & -- \\

2026 Jan 16 07:06 &
\(-1.65\pm0.29\) & \(-1.56\pm0.42\) &
\(-3.25^{+0.09}_{-0.35}\) &
-- & -- \\

2026 Jan 18 07:23 &
\(-1.78\pm0.37\) & \(-1.32\pm0.45\) &
\(-3.25^{+0.09}_{-0.36}\) &
-- & -- \\
\hline
\end{tabular}

\vspace{2pt}
\begin{flushleft}
\footnotesize
Notes. All slopes were measured or fitted over
\(4\times10^4 < \rho < 8\times10^4~{\rm km}\), where
\(S(\rho)\propto\rho^k\).
The quantity \(k_{\rm azi}\) is measured from the full azimuthally
averaged profile, while \(k_{\perp}\) is measured from the profile
perpendicular to the projected Sun--comet axis.
The quantity \(k_{\rm proj}\) is predicted by the empirical
fixed-morphology projection model. The Haser-mixture parameters are derived from the azimuthally averaged profiles and use the average January profile as the late-time baseline.
The parameter \(\eta\) is the relative weight of the fading component,
and \(\ell_{\rm DE}\) is the projected destruction length.
Uncertainties are the 16th--84th percentile ranges from Monte Carlo
resampling.
\(^{\dagger}\)The fading component is weak and
\(\ell_{\rm DE}\) is poorly constrained.
\end{flushleft}
\end{table*}

The fits show that the projected destruction length is relatively stable while the fading component is clearly detected.  From UT 2025 November 25 to UT 2025 December 20, we find \(\ell_{\rm DE}\sim(6.4\text{--}7.2)\times10^4~{\rm km}\).  Over the same interval, the fitted amplitude decreases from \(\eta\sim1.9\) to \(\eta\sim0.4\), showing that the post-perihelion evolution is driven mainly by the weakening of the fading component rather than by a large change in its characteristic length. After late December, \(\eta\) becomes small and \(\ell_{\rm DE}\) is poorly constrained. The color profile provides a separate check on this length scale. In the November profile in Fig.~\ref{color}, the blueward color trend begins to level off near \(\rho_{\rm col}\sim6\times10^4~{\rm km}\), which we use as an empirical color-transition scale. This radius is comparable to the Haser-mixture destruction length measured while the fading component is still detected, \(\ell_{\rm DE}\sim(6.4\text{--}7.2)\times10^4~{\rm km}\).


\subsection{Properties of Icy Grains}
\label{Properties of Icy Grains}

The destruction length measured in Sect.~\ref{Deviation of surface brightness} characterizes the radial scale over which the fading component loses optical scattering cross section. A relevant comparison is the survival distance of volatile-bearing icy grains after release. We therefore computed the thermal evolution of compact spherical icy grains containing different amounts of absorbing carbonaceous material. The detailed radiative-equilibrium calculation, sublimation prescription, and size-loss evolution estimate are described in Appendix~\ref{appendix_icy_grain_calculations}.

In the model, water ice is treated as the host material and carbonaceous material as the absorbing inclusion. The effective complex refractive index is computed using the Maxwell--Garnett effective medium approximation \citep{garnett1904xii,aspnes1982local}. We use the revised water-ice optical constants of \citet{warren2008optical} and the carbon optical constants of \citet{edoh1983optical}. The main effect of adding a small absorbing carbonaceous fraction is to raise the grain temperature and therefore increase the sublimation rate by orders of magnitude relative to pure ice.

\begin{figure}
    \centering

    \begin{minipage}{\textwidth}
        \centering
        \includegraphics[width=\textwidth]{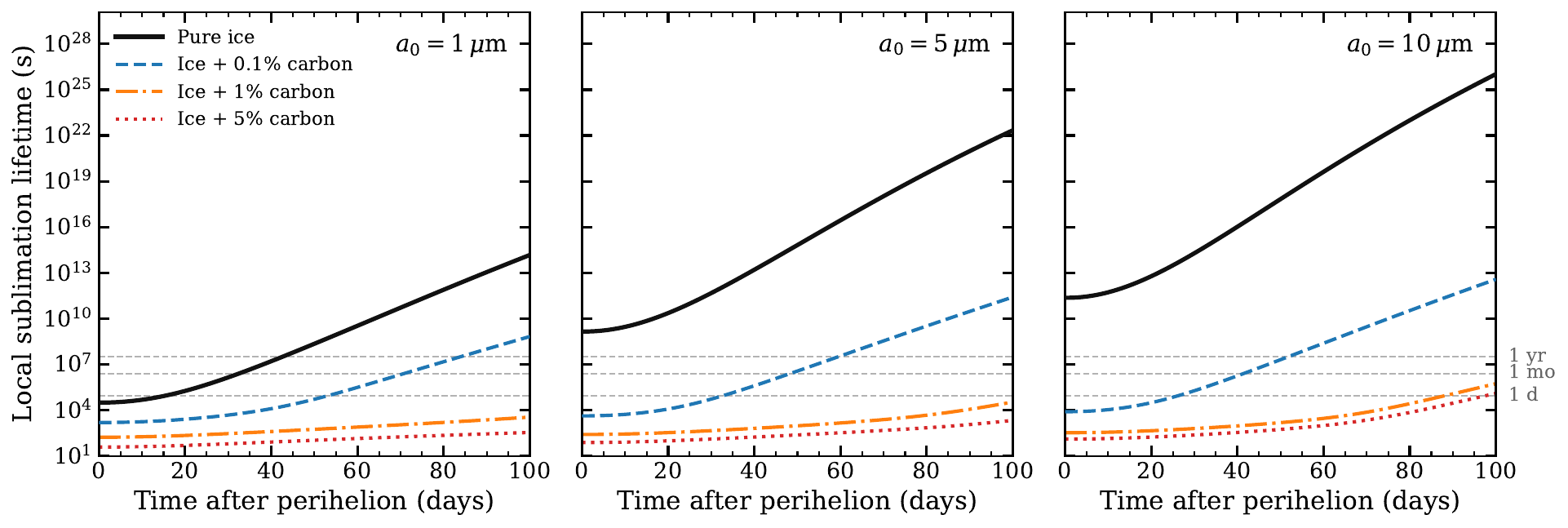}
    \end{minipage}

    \begin{minipage}{\textwidth}
        \centering
        \includegraphics[width=\textwidth]{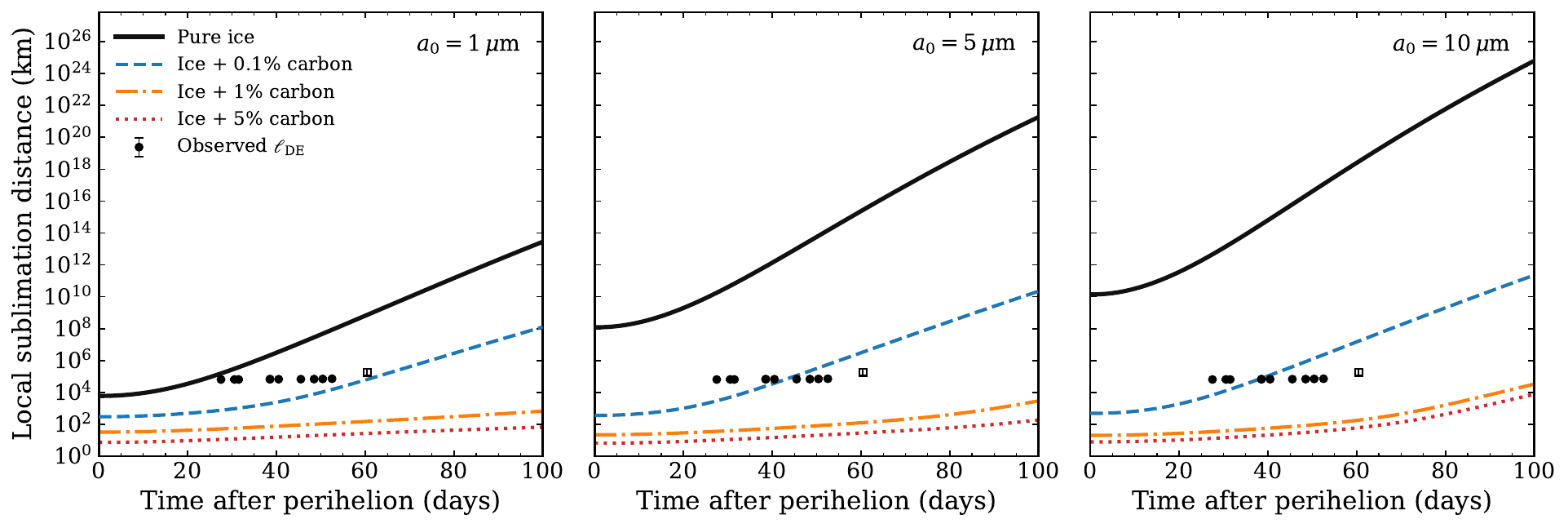}
    \end{minipage}

    \caption{Local sublimation lifetimes and sublimation distances of icy grains. Top: local sublimation timescale \(\tau_{\rm loc}\) for grains with \(a_0=1,5,\) and \(10~\mu{\rm m}\). Horizontal dashed lines mark 1 day, 1 month, and 1 year. Bottom: corresponding sublimation distance. Filled symbols show the destruction-length measurements, while the open symbol marks the measurement with a larger inferred scale and larger uncertainty.}
    \label{fig:LIFEandDIST}
\end{figure}

For a grain with initial radius \(a_0\), we compare the observed destruction length with the local sublimation distance, \(\ell_{\rm sub}=v_{\rm ej}\tau_{\rm loc}\), where \(\tau_{\rm loc}\) is the local sublimation lifetime and \(v_{\rm ej}=190\,\beta^{1/2}~{\rm m~s^{-1}}\) is adopted from Sect.~\ref{Morphology}. Fig.~\ref{fig:LIFEandDIST} shows the resulting local lifetimes and sublimation distances for \(a_0=1,5,\) and \(10~\mu{\rm m}\) grains. Pure ice grains are much too long lived, producing sublimation distances far larger than the measured destruction length. In contrast, grains containing \(1\%\) or \(5\%\) carbon are generally too short lived and yield sublimation distances mostly below the observed scale. The best match is obtained for weakly absorbing carbon-bearing grains. In particular, \(a_0\sim5\)--\(10~\mu{\rm m}\) grains with \(f_{\rm C}=0.1\%\) give local sublimation distances comparable to the observed \(\ell_{\rm DE}\sim(6\text{--}7)\times10^4~{\rm km}\) during the interval when the fading component is clearly detected. Thus, the destruction scale favors micron- to tens-of-micron icy grains containing a small but nonzero absorbing refractory fraction.

The inferred transient icy-grain component does not require icy grains to dominate the overall post-perihelion optical scattering. Post-perihelion polarimetry obtained within aperture radii of approximately $(1.5$--$2.0)\times10^{4}$~km has been interpreted as being dominated by refractory aggregates \citep{choi2026dust}. Notably, the photopolarimetric aperture broadly overlaps the region in which our November color profile remains relatively red, with $g-r=0.63\pm0.01$ over $(1$--$5)\times10^{3}$~km and $g-r=0.74\pm0.01$ over $(1$--$2)\times10^{4}$~km, whereas the color becomes anomalously blue only at larger projected distances, reaching $g-r=0.30\pm0.02$ over $(4$--$8)\times10^{4}$~km. For the $5$--$10~\mu$m grains favored by our sublimation model, outward transport followed by loss of optical cross section naturally produces spatially extended fading on scales comparable to those observed. The polarimetric and photometric constraints can therefore be mutually consistent if the volatile-bearing grains represent a spatially and temporally limited component superposed on a more persistent refractory-dust coma.
 
We also ask whether the same grain model can account for the observed water production. The optical cross section measured from the photometry in Tab.~\ref{tab:cross}, which was derived assuming \(p_r=0.04\), is converted to an equivalent icy-grain cross section by scaling \(C_{\rm ice}=C_{\rm e}(p_r/p_{\rm ice})\). We adopt \(p_{\rm ice}=0.5\), estimated using the approach of \citet{stevenson2012near} for a median projected travel distance of \(\sim6.5\times10^4\) km at \(r_{\rm H}\sim2.3\) au, consistent with weakly carbon-contaminated ice with \(f_{\rm C}=0.1\%\). The corresponding water-production calculation is described in Appendix~\ref{appendix_icy_grain_calculations}.

With no additional normalization, the model gives \(Q_{\rm H_2O,grain}\sim2.1\times10^{28}~{\rm s^{-1}}\) at 27 days after perihelion, decreasing to \(\sim8.7\times10^{27}~{\rm s^{-1}}\) by 31 days and \(\sim9\times10^{26}~{\rm s^{-1}}\) by 40 days. These values are of the correct order of magnitude for the late-November measurements, including the SOHO/SWAN and TRAPPIST-North values. At later times, however, the modeled production rate decreases more rapidly than the large-aperture water-production measurements. This indicates that the fading component should not be interpreted as the complete source of all observed water vapor. Instead, it likely represents one short-lived volatile-bearing grain population, while other sources such as nucleus sublimation, continued release of fresh icy grains, grains with a range of sizes and impurity fractions, and aperture-averaging effects may also contribute.

\subsection{Icy-grain activity in an interstellar context}

The fading component inferred for comet 3I places this object in the broader context of icy-grain activity in comets. In Solar System comets, volatile-bearing grains have been invoked to explain extended volatile production, radial fading of coma structures, and color variations, as illustrated by cases such as 103P/Hartley 2, 67P/Churyumov--Gerasimenko, and 46P/Wirtanen \citep{protopapa2014water,gicquel2016sublimation,protopapa2021nondetection}. Thus, the behavior inferred for comet 3I has Solar System analogues. The contrast with 2I/Borisov, whose dust coma has been interpreted as dominated by compact millimeter-scale pebbles with no strong evidence for grain sublimation or fragmentation \citep{yang2021compact}, further suggests that interstellar comets may exhibit diverse modes of volatile-bearing dust activity.

In comet 3I, the grain population consistent with the observed fading scale provides a physical framework for the transient outer-coma steepening, the disappearance of the blue outer-coma color structure, and the order of magnitude of the water production near the main imaging epoch. Pure water ice survives over distances much larger than the measured destruction length, whereas strongly absorbing icy grains are destroyed too rapidly. The observed scale instead favors an intermediate regime, in which a small absorbing refractory component raises the sublimation efficiency while still allowing survival over projected distances of order \(10^5\) km. This balance between survival and sublimation efficiency is the constraint imposed by the ZTF data. Although the inferred grain properties are model-dependent, the combined brightness, color, and fading-length constraints show that the post-perihelion optical evolution of comet 3I can be linked to the survival of short-lived volatile-bearing grains.

\section{Conclusions}
\label{Conclusions}

We have analyzed ZTF observations of the interstellar comet 3I/ATLAS. The main results are as follows.

\begin{enumerate}
    \item Fixed-aperture photometry shows different pre- and post-perihelion behavior. Before perihelion, the heliocentric index increases with aperture size, from $\gamma=1.63\pm0.14$ in the $5\times10^3$ km aperture to $1.88\pm0.34$ in the $2\times10^4$ km aperture, indicating a steeper inbound brightening in the outer part of the sampled coma. After perihelion, the trend reverses, with $\gamma$ decreasing from $2.44\pm0.09$ in the $5\times10^3$ km aperture to $1.54\pm0.09$ in the $8\times10^4$ km aperture. Thus, the outer coma shows a steeper inbound rise but a shallower outbound decline than the inner coma.
    
    \item Under the adopted grain-size and residence-time assumptions, the photometry implies model-dependent dust mass-loss rates of \(\dot{M}_d\sim123\text{--}203~{\rm kg~s^{-1}}\) before perihelion and \(\sim596\text{--}1412~{\rm kg~s^{-1}}\) after perihelion. Extrapolating the fitted mass-loss laws to perihelion gives \(\dot{M}_{d,{\rm pre}}=849\pm244~{\rm kg~s^{-1}}\) and \(\dot{M}_{d,{\rm post}}=2164\pm84~{\rm kg~s^{-1}}\). Integrating these laws over the observed perihelion passage gives a total modeled dust loss of \(M_{\rm d,tot}\sim1.2\times10^10~{\rm kg}\), corresponding to a dust-equivalent sphere of radius \(\sim180\) m for \(\rho_d=500~{\rm kg~m^{-3}}\).

    \item The projected sunward turnaround distance gives \(v_{\rm ej}\sim(190\pm3)a_{\mu{\rm m}}^{-1/2}~{\rm m~s^{-1}}\). The tail width on 2026 January 18 gives an effective perpendicular velocity of \(v_\perp=(79\pm6)a_{\mu{\rm m}}^{-1/2}~{\rm m~s^{-1}}\). Together, these velocities imply an effective post-perihelion dust-ejection opening angle of \(\sim45\pm3^\circ\), consistent with a broad angular distribution of ejecta.

    \item The early post-perihelion coma showed excess outer-profile steepening, reaching \(k\sim-2.78\), together with an anomalously blue outer \(g-r\) color. Both signatures weakened by January, consistent with a transient fading scattering component.

    \item The fading component is characterized with a destruction length of \(\ell_{\rm DE}\sim(6.4\text{--}7.2)\times10^4\) km. Thermophysical sublimation calculations show that the observed destruction scale is best matched by weakly absorbing icy grains with \(a_0\sim5\)--\(10~\mu{\rm m}\) and \(f_{\rm C}\sim0.1\%\).

    \item The derived lightly contaminated icy-grain population can supply the correct order of magnitude of water production near the main imaging epoch, but it does not reproduce the later large-aperture water-production measurements. This suggests that the fading component represents one short-lived volatile-bearing grain population rather than the complete source of the observed water vapor.
\end{enumerate}

\begin{acknowledgments}
The work is supported by the Science and Technology Development Fund (FDCT) of Macau (grant Nos. 0008/2024/AKP, 0034/2024/AMJ, 0008/2024/ITP1, 002/2024/SKL, 0002/2025/AKP), the National Key Research and Development Program of China (grant No.2024YFE0201000), and the National Natural Science Foundation of China (No.~12472048).
\end{acknowledgments}




\appendix
Appendix~\ref{appendix_auxiliary_figures} presents auxiliary figures
supporting the morphology, mass-loss, and icy-grain analyses, together
with tables listing the apparent and absolute magnitudes corresponding
to the effective scattering cross-sections in
Table~\ref{tab:cross} and the sunward-turnaround measurements. Appendix~\ref{appendix:projection_model}
describes the projection-sensitivity test used to assess the effect of
changing viewing geometry on the measured outer-coma slopes.
Appendix~\ref{appendix_icy_grain_calculations} describes the
thermal-equilibrium, sublimation, and water-production calculations
used in Sect.~\ref{Properties of Icy Grains}.

\section{Supplementary Figures AND TABLES}
\label{appendix_auxiliary_figures}
\setcounter{figure}{0}
\renewcommand{\thefigure}{A\arabic{figure}}
\renewcommand{\theHfigure}{A\arabic{figure}}
\setcounter{table}{0}
\renewcommand{\thetable}{A\arabic{table}}
\renewcommand{\theHtable}{A\arabic{table}}
\noindent 

\begin{figure}
\centering
    \includegraphics[width=0.6\columnwidth]{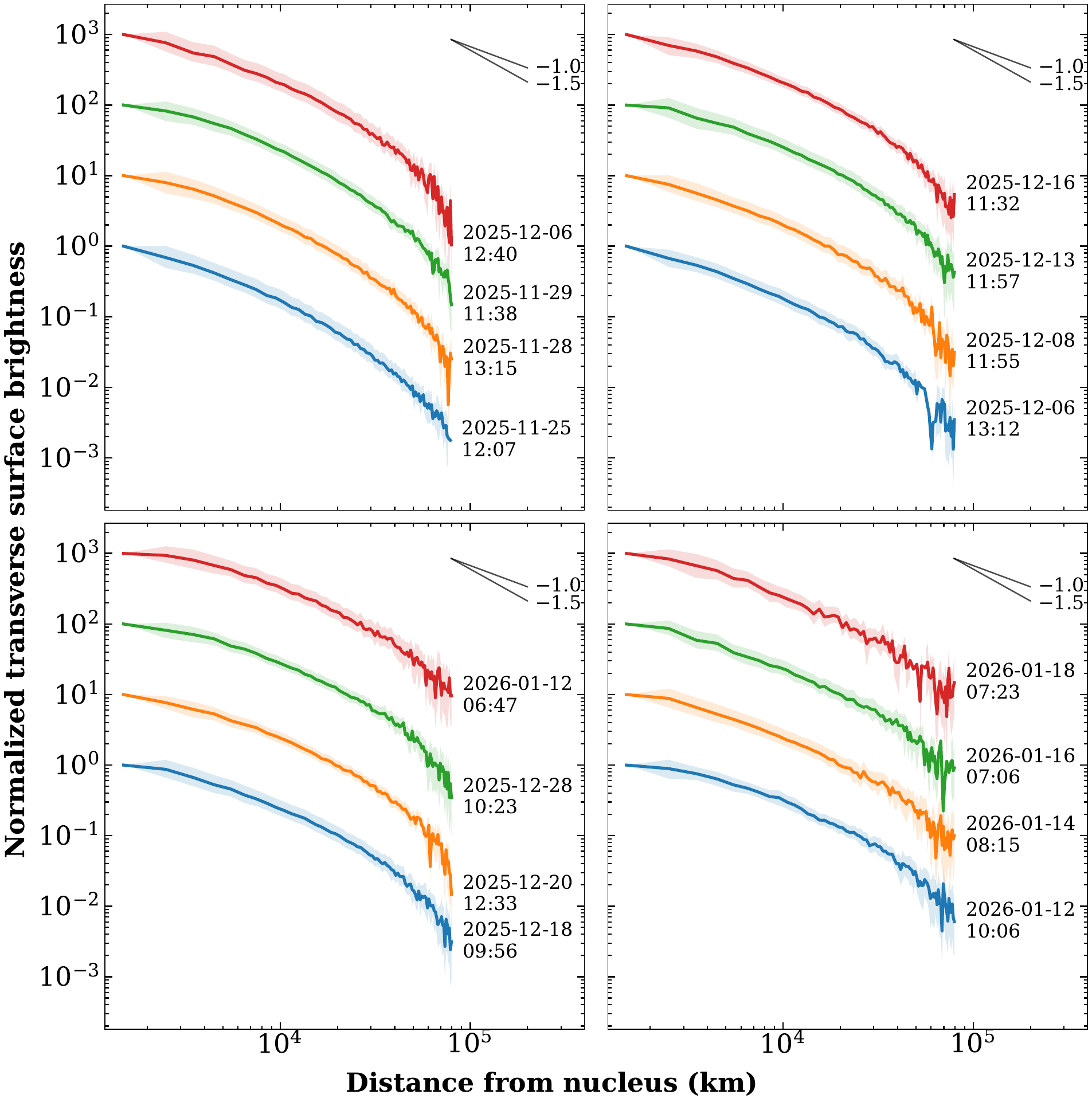}
    \caption{Normalized surface-brightness profiles extracted perpendicular
to the projected Sun--comet axis for the 16 post-perihelion ZTF epochs.
Opposite perpendicular directions were combined. The profiles are
vertically offset for clarity, and the shaded regions represent the
surface-brightness uncertainties. Short reference segments with slopes of -1 and -1.5 are plotted in the upper-right corner of each panel.}
    \label{fig:transverse_profiles}
\end{figure}

\begin{figure}
\centering
    \includegraphics[width=0.5\columnwidth]{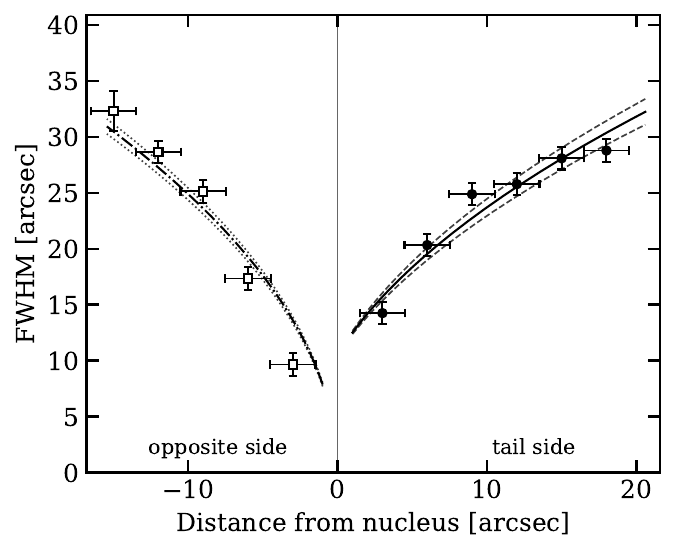}
    \caption{FWHM of the dust tail as a function of distance from the nucleus on 2026 January 18. Positive and negative distances denote the tail and anti-tail sides, respectively. Horizontal error bars indicate the segments along the projected tail axis over which the profiles were measured, and the vertical error bars show the FWHM uncertainties. Filled circles and open squares mark the tail and anti-tail measurements. The solid and dash-dotted curves are separate fits to Eq.~\ref{eq:tail_width_final}, and the dashed/dotted curves show the corresponding $\pm1\sigma$ velocity ranges. The vertical line marks the nucleus.}
    \label{FWHM_two_sides_signed}
\end{figure}

\begin{figure}
\centering
    \includegraphics[width=0.5\columnwidth]{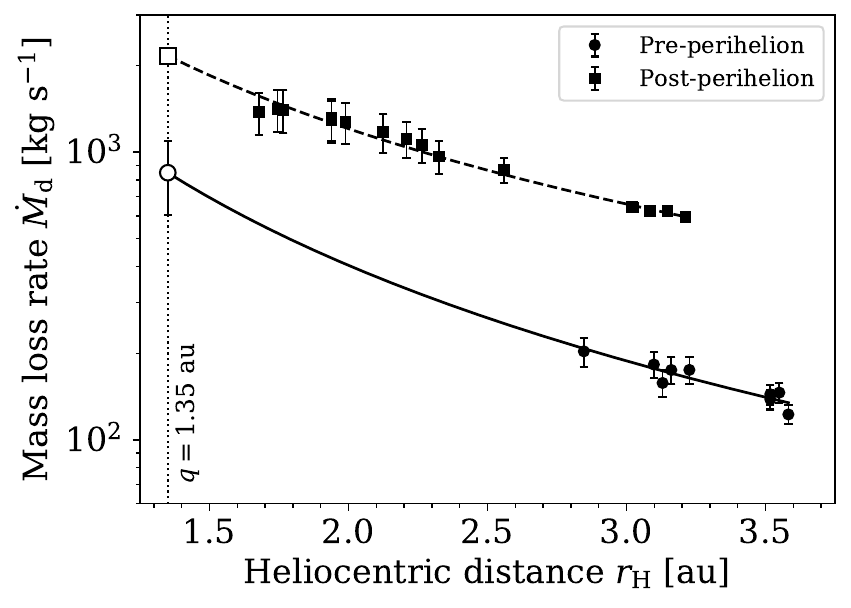}
    \caption{Dust mass-loss rate as a function of heliocentric distance. Filled circles and filled squares show the pre- and post-perihelion measurements, respectively, derived from the largest photometric apertures. Error bars represent the uncertainties. The solid and dashed curves show separate power-law fits, giving \(\dot{M}_d\propto r_H^{-1.88\pm0.34}\) before perihelion and \(\dot{M}_d\propto r_H^{-1.54\pm0.09}\) after perihelion. The vertical dotted line marks the perihelion distance, \(q=1.35\) au, and the open symbols show the corresponding extrapolated mass-loss rates at perihelion.}
    \label{dustloss}
\end{figure}

\begin{figure}
    \centering
    \includegraphics[width=0.9\linewidth]{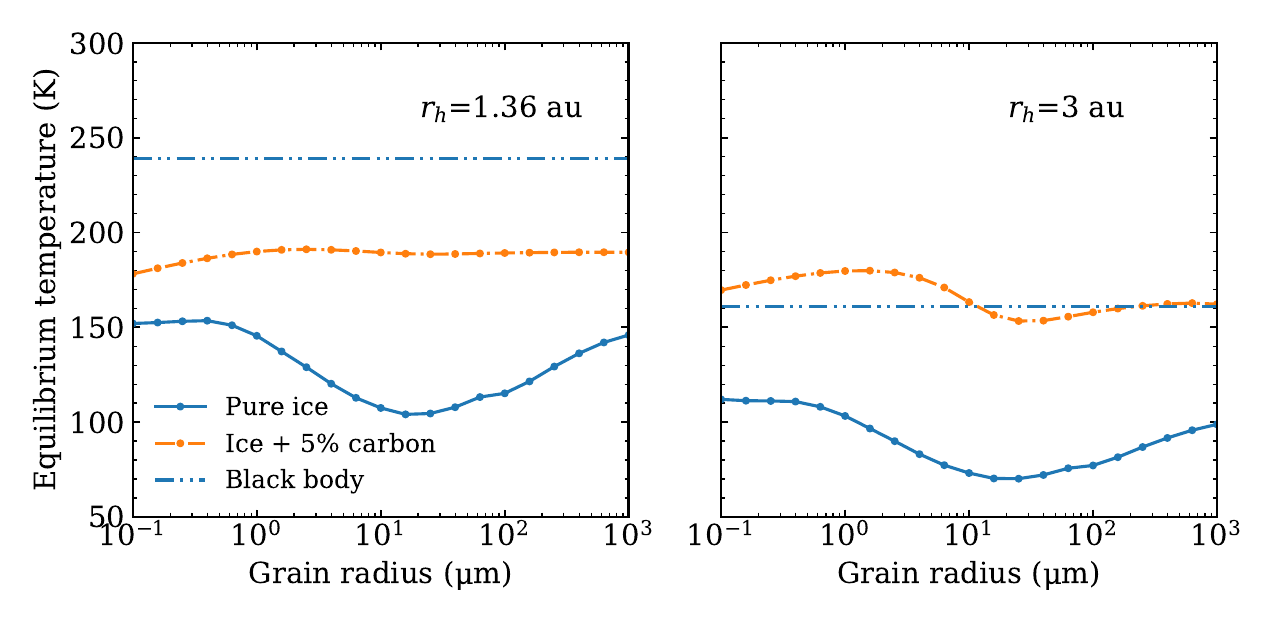}
    \caption{
    Equilibrium temperature of compact spherical icy grains as a function of
    grain radius. The left and right panels show results at
    \(r_{\rm H}=1.36\) au and \(r_{\rm H}=3\) au, respectively. Solid blue
    curves show pure water ice, while orange curves show water ice containing
    carbonaceous inclusions with volume fraction \(f_{\rm C}=5\%\). The
    horizontal dash-dotted lines show the corresponding blackbody temperatures.
    }
    \label{fig:equilibrium_temperature}
\end{figure}

\begin{figure}
\centering

\begin{minipage}{0.49\textwidth}
    \centering
    \includegraphics[width=\textwidth]{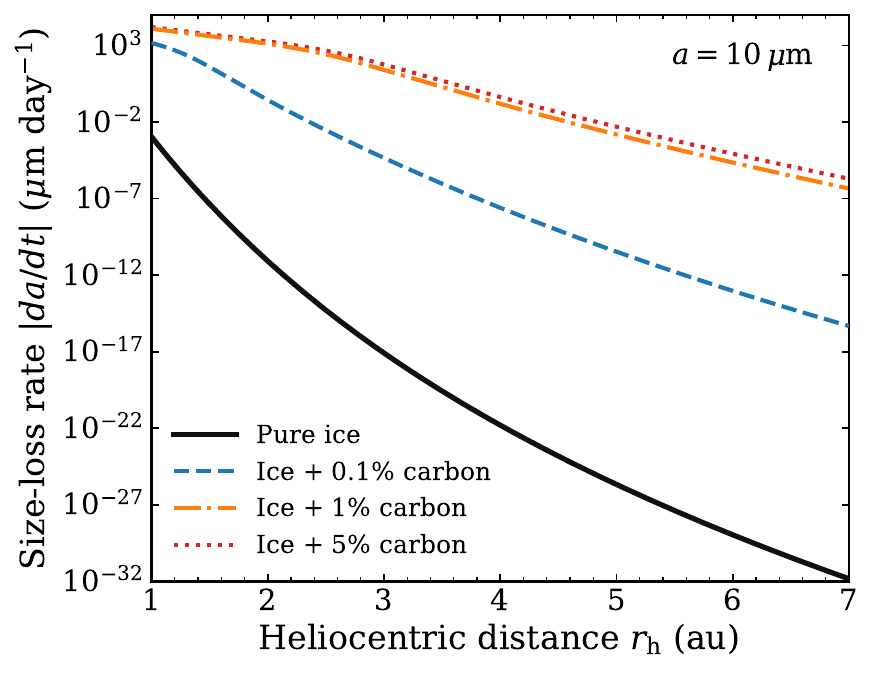}
\end{minipage}
\hfill
\begin{minipage}{0.49\textwidth}
    \centering
    \includegraphics[width=0.95\textwidth]{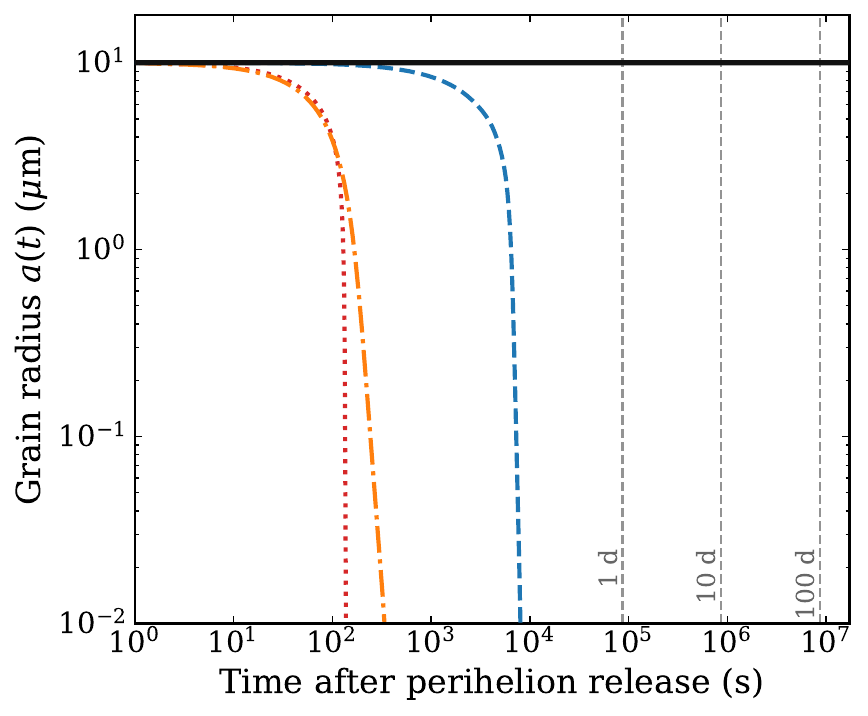}
\end{minipage}

\caption{Left: size-loss rate \(\left|da/dt\right|\) as a function of heliocentric distance for \(10~\mu{\rm m}\) pure water-ice grains and carbon-bearing icy grains. Right: radius evolution of \(10~\mu{\rm m}\) grains released at the perihelion of comet 3I. Vertical dashed lines mark 1, 10, and 100 days after release.}
\label{fig:size_loss_and_evolution}
\end{figure}

\begin{figure}
    \centering
    \includegraphics[width=0.8\linewidth]{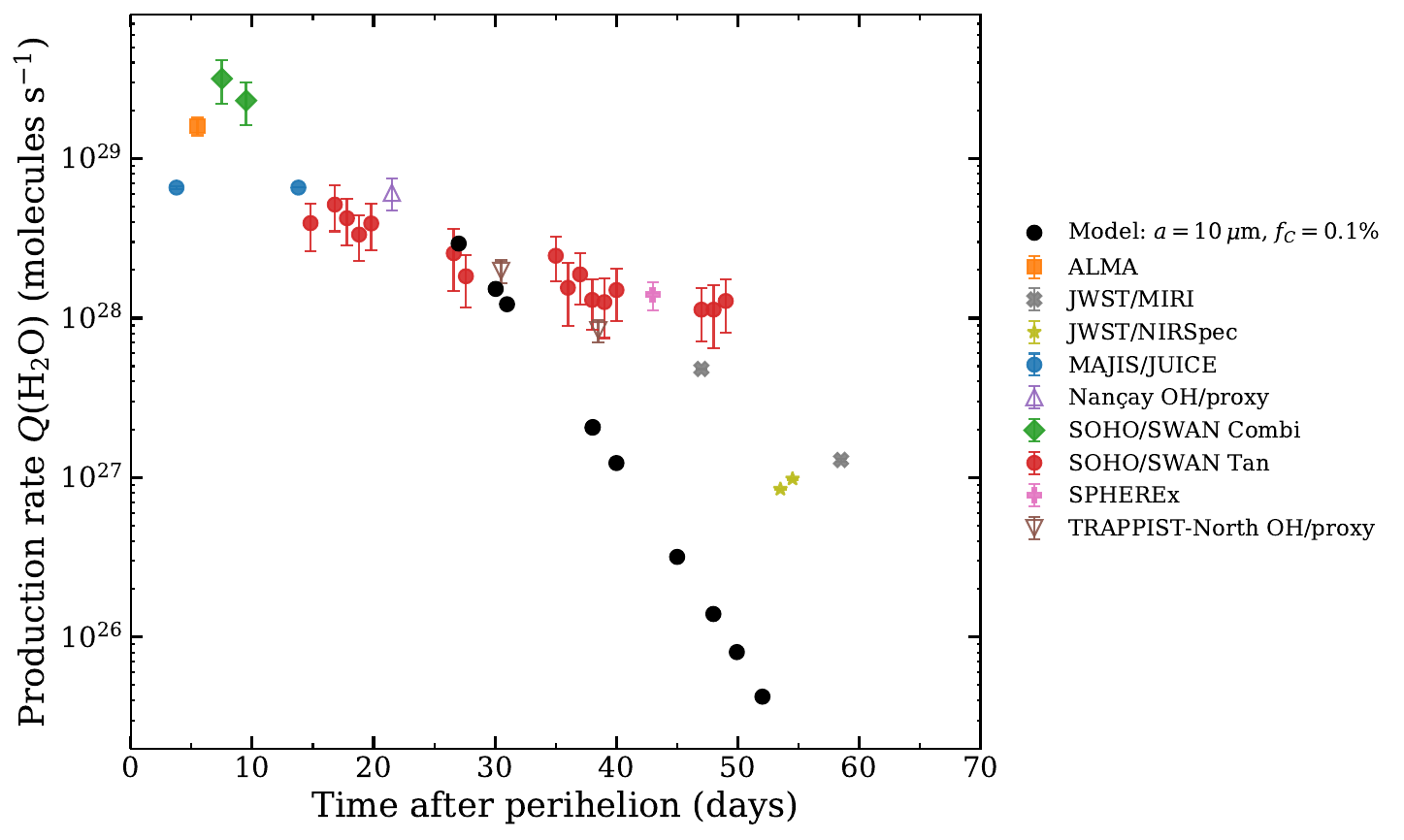}
    \caption{Comparison between observed water production rates and the water production estimated from the optical scattering cross section. The black points show the model for \(a=10~\mu{\rm m}\), \(f_{\rm C}=0.1\%\), and an assumed icy-grain albedo \(p_{\rm ice}=0.5\). Colored symbols show observed H\(_2\)O production rates, with OH-based values shown as proxies where applicable \citep{bockelee2026detection,salazar2026water,combi2026water,tan2026perihelion,biver2026perihelion,lisse2026spherex,belyakov2026volatile,roth2026isotopic}.}
    \label{Water_Production_Rates}
\end{figure}

\begin{table}
\centering
\renewcommand{\arraystretch}{1.15}

\caption{Apparent magnitudes measured in the ZTF $r$ band within projected circular apertures of radius $\rho$.}
\label{tab_app}
\setlength{\tabcolsep}{5pt}

\textbf{Panel A. Pre-perihelion photometry}

\vspace{3pt}
\begin{tabular}{lccc}
    \toprule
    & \multicolumn{3}{c}{$m_r$} \\
    \cmidrule(lr){2-4}
    UT Date
    & $\rho=5\times10^{3}$ km
    & $\rho=10\times10^{3}$ km
    & $\rho=20\times10^{3}$ km \\
    \midrule
    29 Jul 2025 05:03 & $17.47 \pm 0.02$ & $16.93 \pm 0.02$ & $16.81 \pm 0.03$ \\
    30 Jul 2025 05:04 & $17.40 \pm 0.02$ & $16.92 \pm 0.02$ & $16.60 \pm 0.03$ \\
    31 Jul 2025 05:03 & $17.31 \pm 0.02$ & $16.78 \pm 0.02$ & $16.58 \pm 0.02$ \\
    31 Jul 2025 05:07 & $17.32 \pm 0.02$ & $16.80 \pm 0.02$ & $16.60 \pm 0.02$ \\
    09 Aug 2025 05:06 & $16.98 \pm 0.04$ & $16.41 \pm 0.04$ & $16.20 \pm 0.04$ \\
    11 Aug 2025 05:53 & $16.93 \pm 0.03$ & $16.34 \pm 0.03$ & $16.14 \pm 0.04$ \\
    12 Aug 2025 05:28 & $16.81 \pm 0.02$ & $16.50 \pm 0.03$ & $16.25 \pm 0.03$ \\
    13 Aug 2025 05:03 & $16.87 \pm 0.02$ & $16.30 \pm 0.02$ & $16.07 \pm 0.02$ \\
    21 Aug 2025 04:37 & $16.57 \pm 0.03$ & $15.99 \pm 0.03$ & $15.75 \pm 0.03$ \\
    \bottomrule
\end{tabular}

\vspace{8pt}

\textbf{Panel B. Post-perihelion photometry}

\vspace{3pt}
\begin{tabular}{lccccc}
    \toprule
    & \multicolumn{5}{c}{$m_r$} \\
    \cmidrule(lr){2-6}
    UT Date
    & $\rho=5\times10^{3}$ km
    & $\rho=10\times10^{3}$ km
    & $\rho=20\times10^{3}$ km
    & $\rho=40\times10^{3}$ km
    & $\rho=80\times10^{3}$ km \\
    \midrule
    25 Nov 2025 12:07 & $13.08 \pm 0.03$ & $12.18 \pm 0.03$ & $11.50 \pm 0.03$ & $10.96 \pm 0.03$ & $10.57 \pm 0.03$ \\
    28 Nov 2025 13:15 & $13.22 \pm 0.03$ & $12.32 \pm 0.03$ & $11.61 \pm 0.03$ & $11.05 \pm 0.03$ & $10.65 \pm 0.03$ \\
    29 Nov 2025 11:38 & $13.30 \pm 0.03$ & $12.38 \pm 0.03$ & $11.66 \pm 0.03$ & $11.09 \pm 0.03$ & $10.68 \pm 0.03$ \\
    06 Dec 2025 12:40 & $13.62 \pm 0.03$ & $12.68 \pm 0.03$ & $11.93 \pm 0.03$ & $11.32 \pm 0.03$ & $10.88 \pm 0.03$ \\
    06 Dec 2025 13:12 & $13.59 \pm 0.03$ & $12.66 \pm 0.03$ & $11.91 \pm 0.03$ & $11.30 \pm 0.03$ & $10.87 \pm 0.03$ \\
    08 Dec 2025 11:55 & $13.69 \pm 0.03$ & $12.76 \pm 0.03$ & $12.00 \pm 0.03$ & $11.38 \pm 0.03$ & $10.94 \pm 0.03$ \\
    13 Dec 2025 11:57 & $13.91 \pm 0.03$ & $12.98 \pm 0.03$ & $12.20 \pm 0.03$ & $11.55 \pm 0.03$ & $11.06 \pm 0.03$ \\
    16 Dec 2025 11:32 & $14.06 \pm 0.03$ & $13.12 \pm 0.03$ & $12.34 \pm 0.03$ & $11.66 \pm 0.03$ & $11.15 \pm 0.03$ \\
    18 Dec 2025 09:56 & $14.16 \pm 0.03$ & $13.22 \pm 0.03$ & $12.44 \pm 0.03$ & $11.75 \pm 0.03$ & $11.23 \pm 0.03$ \\
    20 Dec 2025 12:33 & $14.27 \pm 0.03$ & $13.31 \pm 0.03$ & $12.52 \pm 0.03$ & $11.82 \pm 0.03$ & $11.15 \pm 0.03$ \\
    28 Dec 2025 10:23 & $14.69 \pm 0.03$ & $13.67 \pm 0.03$ & $12.84 \pm 0.03$ & $12.14 \pm 0.03$ & $11.55 \pm 0.03$ \\
    12 Jan 2026 06:47 & $15.54 \pm 0.03$ & $14.52 \pm 0.03$ & $13.68 \pm 0.03$ & $12.90 \pm 0.03$ & $12.16 \pm 0.03$ \\
    12 Jan 2026 10:06 & $15.46 \pm 0.03$ & $14.50 \pm 0.03$ & $13.68 \pm 0.03$ & $12.91 \pm 0.03$ & $12.25 \pm 0.03$ \\
    14 Jan 2026 08:15 & $15.53 \pm 0.02$ & $14.59 \pm 0.02$ & $13.78 \pm 0.02$ & $13.01 \pm 0.02$ & $12.33 \pm 0.02$ \\
    16 Jan 2026 07:06 & $15.62 \pm 0.03$ & $14.69 \pm 0.02$ & $13.88 \pm 0.02$ & $13.11 \pm 0.02$ & $12.34 \pm 0.02$ \\
    18 Jan 2026 07:23 & $15.72 \pm 0.04$ & $14.78 \pm 0.03$ & $13.97 \pm 0.03$ & $13.20 \pm 0.03$ & $12.55 \pm 0.03$ \\
    \bottomrule
\end{tabular}

\vspace{2pt}
\begin{flushleft}
\footnotesize
\textit{Note.} Here, $\rho$ is the projected radius of the circular photometric aperture at the comet. The uncertainties include the photometric measurement uncertainty and the uncertainty in the photometric zero point.
\end{flushleft}

\end{table}

\begin{table}
\centering
\renewcommand{\arraystretch}{1.15}

\caption{Absolute magnitudes derived from the ZTF $r$-band photometry within projected circular apertures of radius $\rho$.}
\label{tab_abs}
\setlength{\tabcolsep}{5pt}

\textbf{Panel A. Pre-perihelion photometry}

\vspace{3pt}
\begin{tabular}{lccc}
    \toprule
    & \multicolumn{3}{c}{$H_r$} \\
    \cmidrule(lr){2-4}
    UT Date
    & $\rho=5\times10^{3}$ km
    & $\rho=10\times10^{3}$ km
    & $\rho=20\times10^{3}$ km \\
    \midrule
    29 Jul 2025 05:03 & $12.09 \pm 0.08$ & $11.56 \pm 0.08$ & $11.44 \pm 0.08$ \\
    30 Jul 2025 05:04 & $12.04 \pm 0.08$ & $11.57 \pm 0.08$ & $11.24 \pm 0.08$ \\
    31 Jul 2025 05:03 & $11.98 \pm 0.08$ & $11.45 \pm 0.08$ & $11.25 \pm 0.08$ \\
    31 Jul 2025 05:07 & $11.99 \pm 0.08$ & $11.47 \pm 0.08$ & $11.27 \pm 0.08$ \\
    09 Aug 2025 05:06 & $11.83 \pm 0.11$ & $11.25 \pm 0.11$ & $11.05 \pm 0.11$ \\
    11 Aug 2025 05:53 & $11.82 \pm 0.11$ & $11.23 \pm 0.11$ & $11.03 \pm 0.11$ \\
    12 Aug 2025 05:28 & $11.79 \pm 0.11$ & $11.21 \pm 0.11$ & $10.98 \pm 0.11$ \\
    13 Aug 2025 05:03 & $11.80 \pm 0.11$ & $11.23 \pm 0.11$ & $11.00 \pm 0.11$ \\
    21 Aug 2025 04:37 & $11.67 \pm 0.13$ & $11.09 \pm 0.13$ & $10.84 \pm 0.13$ \\
    \bottomrule
\end{tabular}

\vspace{8pt}

\textbf{Panel B. Post-perihelion photometry}

\vspace{3pt}
\begin{tabular}{lccccc}
    \toprule
    & \multicolumn{5}{c}{$H_r$} \\
    \cmidrule(lr){2-6}
    UT Date
    & $\rho=5\times10^{3}$ km
    & $\rho=10\times10^{3}$ km
    & $\rho=20\times10^{3}$ km
    & $\rho=40\times10^{3}$ km
    & $\rho=80\times10^{3}$ km \\
    \midrule
    25 Nov 2025 12:07 & $9.70 \pm 0.18$ & $8.80 \pm 0.18$ & $8.11 \pm 0.18$ & $7.58 \pm 0.18$ & $7.19 \pm 0.18$ \\
    28 Nov 2025 13:15 & $9.78 \pm 0.19$ & $8.88 \pm 0.19$ & $8.17 \pm 0.19$ & $7.61 \pm 0.19$ & $7.21 \pm 0.19$ \\
    29 Nov 2025 11:38 & $9.85 \pm 0.19$ & $8.92 \pm 0.19$ & $8.20 \pm 0.19$ & $7.64 \pm 0.19$ & $7.22 \pm 0.19$ \\
    06 Dec 2025 12:40 & $10.05 \pm 0.18$ & $9.12 \pm 0.18$ & $8.37 \pm 0.18$ & $7.76 \pm 0.18$ & $7.32 \pm 0.18$ \\
    06 Dec 2025 13:12 & $10.02 \pm 0.18$ & $9.10 \pm 0.18$ & $8.35 \pm 0.18$ & $7.74 \pm 0.18$ & $7.30 \pm 0.18$ \\
    08 Dec 2025 11:55 & $10.10 \pm 0.18$ & $9.17 \pm 0.18$ & $8.42 \pm 0.18$ & $7.79 \pm 0.18$ & $7.35 \pm 0.18$ \\
    13 Dec 2025 11:57 & $10.27 \pm 0.17$ & $9.34 \pm 0.17$ & $8.56 \pm 0.17$ & $7.91 \pm 0.17$ & $7.42 \pm 0.17$ \\
    16 Dec 2025 11:32 & $10.39 \pm 0.16$ & $9.45 \pm 0.16$ & $8.67 \pm 0.16$ & $7.99 \pm 0.16$ & $7.47 \pm 0.16$ \\
    18 Dec 2025 09:56 & $10.47 \pm 0.15$ & $9.53 \pm 0.15$ & $8.75 \pm 0.15$ & $8.06 \pm 0.15$ & $7.54 \pm 0.15$ \\
    20 Dec 2025 12:33 & $10.55 \pm 0.14$ & $9.59 \pm 0.14$ & $8.80 \pm 0.14$ & $8.11 \pm 0.14$ & $7.43 \pm 0.14$ \\
    28 Dec 2025 10:23 & $10.87 \pm 0.11$ & $9.84 \pm 0.11$ & $9.02 \pm 0.11$ & $8.31 \pm 0.11$ & $7.73 \pm 0.11$ \\
    12 Jan 2026 06:47 & $11.39 \pm 0.05$ & $10.37 \pm 0.05$ & $9.52 \pm 0.05$ & $8.75 \pm 0.05$ & $8.00 \pm 0.05$ \\
    12 Jan 2026 10:06 & $11.30 \pm 0.05$ & $10.34 \pm 0.05$ & $9.52 \pm 0.05$ & $8.76 \pm 0.05$ & $8.09 \pm 0.05$ \\
    14 Jan 2026 08:15 & $11.32 \pm 0.04$ & $10.38 \pm 0.04$ & $9.56 \pm 0.04$ & $8.79 \pm 0.04$ & $8.11 \pm 0.04$ \\
    16 Jan 2026 07:06 & $11.34 \pm 0.03$ & $10.41 \pm 0.03$ & $9.60 \pm 0.03$ & $8.83 \pm 0.03$ & $8.06 \pm 0.03$ \\
    18 Jan 2026 07:23 & $11.38 \pm 0.04$ & $10.43 \pm 0.04$ & $9.63 \pm 0.04$ & $8.86 \pm 0.04$ & $8.20 \pm 0.04$ \\
    \bottomrule
\end{tabular}

\vspace{2pt}
\begin{flushleft}
\footnotesize
\textit{Note.} Here, $\rho$ is the projected radius of the circular photometric aperture at the comet. The absolute magnitudes were calculated as
$H_r=m_r-5\log_{10}(r_{\rm H}\Delta)-\zeta\alpha$,
where $r_{\rm H}$ and $\Delta$ are the heliocentric and geocentric distances in au, respectively, $\alpha$ is the phase angle, and $\zeta=0.026\pm0.006$ mag deg$^{-1}$ is the adopted linear phase coefficient.
\end{flushleft}

\end{table}

\begin{table}
\centering

\caption{Projected sunward extents and inferred dust ejection speeds.}
\label{tab:sunward_extent}
\begin{tabular}{lcc}
\hline\hline
UT Date &
$R'_{\rm ta}$ &
$v_{\rm ej}/\beta^{1/2}$ \\
&
(km) &
(m s$^{-1}$) \\
\hline
2025 Jul 29 05:03 & $9042 \pm 407$  & $196.5 \pm 4.4$ \\
2025 Jul 30 05:04 & $8298 \pm 244$  & $186.9 \pm 2.8$ \\
2025 Jul 31 05:03 & $8709 \pm 148$  & $190.2 \pm 1.6$ \\
2025 Jul 31 05:07 & $7875 \pm 133$  & $180.8 \pm 1.5$ \\
2025 Aug 09 05:06 & $9295 \pm 347$  & $191.3 \pm 3.6$ \\
2025 Aug 11 05:53 & $8143 \pm 148$  & $179.3 \pm 1.6$ \\
2025 Aug 12 05:28 & $8944 \pm 129$ & $189.9 \pm 1.5$ \\
2025 Aug 13 05:03 & $8762 \pm 127$  & $186.2 \pm 1.4$ \\
2025 Aug 21 04:37 & $8775 \pm 153$  & $191.1 \pm 1.7$ \\
\hline
\end{tabular}

\begin{flushleft}
\footnotesize
\textit{Note.} $R'_{\rm ta}$ is the projected distance from the
photometric center to the point where the brightness decreases to
10\% of its peak value. Its uncertainty is derived from the
surface-brightness uncertainty. The corresponding uncertainties are
propagated to $v_{\rm ej}$.
\end{flushleft}
\end{table}

\clearpage

\section{Projection Sensitivity Test}
\label{appendix:projection_model}

\setcounter{equation}{0}
\renewcommand{\theequation}{B\arabic{equation}}
\begingroup

An empirical two-dimensional model was constructed to estimate how the
changing viewing geometry could affect the measured outer-coma slope
while keeping the intrinsic coma morphology fixed. The UT 2025
November 25 observation, obtained at phase angle
\(\alpha_0=29.96^\circ\), was adopted as the reference epoch. The
reference image was radially calibrated so that its azimuthally
averaged profile reproduced the measured surface-brightness profile,
thereby retaining the observed angular morphology while imposing the
observed radial distribution.

The reference morphology was expressed in polar coordinates aligned
with the projected Sun--comet axis. At each projected cometocentric
distance, the angular modulation \(M(\rho,\phi)\) was measured in
\(15^\circ\) bins and normalized to have an azimuthal mean of unity.
We separated this modulation into an approximately symmetric floor
\(F_p(\rho)\), defined by the \(p\)th percentile of
\(M(\rho,\phi)\), and a positive asymmetric component
\(A_p(\rho,\phi)=\max[M(\rho,\phi)-F_p(\rho),0]\). The two components
were renormalized so that
\(F_p+\langle A_p\rangle_\phi=1\) at every radius. This empirical
decomposition is intended only to isolate the angular enhancements
responsible for the visibly asymmetric morphology and does not imply
physically distinct dust components.

The phase-angle dependence was introduced by assuming that the
projected extent of the Sun-aligned asymmetric structure scales as
\(\sin\alpha\). For epoch \(i\), we defined
\(s_i=\sin\alpha_i/\sin\alpha_0\), where \(\alpha_i\) is the
corresponding phase angle. If \(u\) and \(v\) denote coordinates
parallel and perpendicular to the projected Sun--comet axis,
respectively, the asymmetric component was sampled at
\(u'=u/s_i\) and \(v'=v\). Therefore, as the phase angle decreases
and \(s_i<1\), the structure is compressed toward the photometric center of the comet along the Sun--comet axis, while its perpendicular extent is unchanged.

The modeled surface-brightness distribution was written as

\begin{equation}
    I_i(u,v)=P_0(\rho)F_p(\rho)
+s_i^{-q}P_0(\rho')
 A_p(\rho',\phi'),
    \label{eq:proje_text}
\end{equation}

where \(P_0(\rho)\) is the measured azimuthally averaged profile of the
reference epoch,
\(\rho'=[(u/s_i)^2+v^2]^{1/2}\), and
\(\phi'=\tan^{-1}[v/(u/s_i)]\). The first term represents the
approximately symmetric component, which was held fixed, whereas the
second term represents the projected asymmetric component. The factor
\(s_i^{-q}\) describes the associated surface-brightness scaling. The
baseline model adopted \(q=1\), corresponding approximately to flux
conservation under one-dimensional compression, while \(q=0\) was also
tested as a case without flux rescaling.

For each epoch, \(I_i\) was azimuthally averaged to obtain the modeled
radial profile, and the projection-model slope \(k_{\rm proj}\) was
fitted over \(4\times10^4<\rho<8\times10^4~{\rm km}\), identical to
the interval used for the observed profiles. The modeled sequence was
anchored to the measured slope of the reference epoch to remove the
small numerical offset introduced by angular resampling. The baseline
calculation used \(p=20\) and \(q=1\). To evaluate the sensitivity to
the empirical decomposition and flux scaling, we repeated the
calculation for \(p=10\), 20, and 30 and for \(q=0\) and 1. The full
range of the resulting slopes was adopted as the model-sensitivity
interval. The resulting values of \(k_{\rm proj}\) are listed in
Table~\ref{tab:haser_mixture_slopes}.

The model predicts progressive steepening of the azimuthally averaged
profile as the phase angle decreases, opposite to the observed
flattening of both the azimuthally averaged and perpendicular profiles.
Changing viewing geometry alone therefore cannot reproduce either the
direction or the magnitude of the measured slope evolution. By
representing the change in viewing geometry as a one-dimensional
compression of the observed two-dimensional morphology along the
projected Sun--comet axis, the model neglects full three-dimensional
line-of-sight integration and possible spatial variations in the dust
scattering properties. Because the intrinsic morphology is held fixed
and three-dimensional dust dynamics are omitted, the model should be
interpreted strictly as a projection-sensitivity test rather than a
physical reconstruction of the coma.
\endgroup
\clearpage
\section{Icy-grain thermal and water-production calculations}
\label{appendix_icy_grain_calculations}

\setcounter{equation}{0}
\renewcommand{\theequation}{C\arabic{equation}}

For a compact spherical grain of radius \(a\) at heliocentric distance \(r_{\rm H}\), the equilibrium temperature is obtained from radiative balance,
\begin{equation}
    \frac{1}{r_{\rm H}^{2}}
    \int_0^\infty
    Q_{\rm abs}(a,\lambda,m_{\rm eff})
    F_{\odot,\lambda}\,d\lambda
    =
    4
    \int_0^\infty
    Q_{\rm abs}(a,\lambda,m_{\rm eff})
    \pi B_{\lambda}(T)\,d\lambda
    +
    4 L_{\rm sub}(T) Z(T) ,
    \label{eq:thermal_balance}
\end{equation}
where \(F_{\odot,\lambda}\) is the solar flux at 1 au, \(Q_{\rm abs}\) is the Mie absorption efficiency, \(B_{\lambda}(T)\) is the Planck function, and \(L_{\rm sub}\) is the latent heat of water-ice sublimation. The sublimation mass flux is
\begin{equation}
    Z(T)
    =
    P_{\rm v}(T)
    \left(
        \frac{m_{\rm H_2O}}{2\pi kT}
    \right)^{1/2},
    \label{eq:sublimation_flux}
\end{equation}
where \(P_{\rm v}(T)\) is the equilibrium vapor pressure of water ice, \(m_{\rm H_2O}\) is the mass of one water molecule, and \(k\) is the Boltzmann constant.

The effective complex refractive index \(m_{\rm eff}(\lambda)\) is computed using the Maxwell--Garnett effective medium approximation \citep{garnett1904xii,aspnes1982local}. Water ice is taken as the host material, while a small volume fraction of refractory absorbing inclusions is embedded in the ice. We adopt carbonaceous material as the refractory absorber and denote its volume fraction by \(f_{\rm C}\). The constituent complex refractive indices are written as \(m_j(\lambda)=n_j(\lambda)+ik_j(\lambda)\). We use the revised water-ice optical constants of \citet{warren2008optical} for \(m_{\rm ice}\), and the carbon optical constants of \citet{edoh1983optical} for the absorbing inclusions. These tabulated constants are interpolated onto a common logarithmic wavelength grid before being combined to obtain \(m_{\rm eff}(\lambda)\).

The resulting equilibrium temperatures are shown in Fig.~\ref{fig:equilibrium_temperature}. Pure ice grains are substantially colder than blackbody grains because they absorb inefficiently at solar wavelengths while still emitting thermally in the infrared. The addition of absorbing carbonaceous inclusions raises the grain temperature over a broad range of sizes. This effect is especially important for micron- to tens-of-micron-sized grains, for which even a small change in temperature leads to a large change in the water-ice sublimation rate.

The equilibrium temperature is converted to a sublimation-driven radius-loss rate using the vapor-pressure parameterization of \citet{mukai1989icy}, as adopted by \citet{moreno2018icy}. We write
\begin{equation}
    \left|\frac{da}{dt}\right|
    =
    1.97\times10^{9}
    T^{7/2}
    \exp\left[
        -\frac{
        5505+165(T/140)^{2.4}
        }{T}
    \right],
    \label{eq:adot_loss}
\end{equation}
where \(T\) is in kelvin and \(|da/dt|\) is in \(\mu{\rm m~day^{-1}}\). The corresponding size-loss rates and radius evolution are shown in Fig.~\ref{fig:size_loss_and_evolution}. Pure water ice has an extremely small size-loss rate over the relevant range of \(r_{\rm H}\). In contrast, even a small amount of carbonaceous material increases the absorption efficiency and raises the grain temperature, producing orders-of-magnitude larger sublimation rates. The strong separation between the pure-ice and carbon-bearing cases reflects the exponential dependence of Eq.~\ref{eq:adot_loss} on \(T\). In the radius-evolution calculation, carbon-bearing \(10~\mu{\rm m}\) grains are eroded to the adopted cutoff radius, \(a_{\rm cut}=10^{-2}~\mu{\rm m}\), in less than one day, whereas pure ice grains remain essentially unchanged over the same interval. Grains with higher carbon fractions are destroyed close to the release time, while more weakly absorbing grains survive longer. This illustrates that the observed destruction scale is highly sensitive to the small absorbing fraction mixed into the ice.

For a grain with initial radius \(a_0\), the local lifetime at heliocentric distance \(r_{\rm H}\) is defined as
\begin{equation}
    \tau_{\rm loc} = \frac{a_0}{\left|da/dt\right|},
    \label{eq:local_lifetime}
\end{equation}
where \(a_0\) is the initial grain radius. The heliocentric distance is fixed to the instantaneous value of comet 3I at each observing epoch, so \(\tau_{\rm loc}\) is a characteristic local sublimation timescale rather than a full orbital lifetime. The corresponding local sublimation distance is
\begin{equation}
    \ell_{\rm sub}
    =
    v_{\rm ej}\tau_{\rm loc},
    \label{eq:local_distance}
\end{equation}
where \(v_{\rm ej}=(190\pm3)\,\beta^{1/2}~{\rm m~s^{-1}}\) is adopted from Sect.~\ref{Morphology}. These quantities are compared with the measured destruction lengths in Fig.~\ref{fig:LIFEandDIST}.

The water production rate from the active icy-grain cross section is estimated from the total sublimating surface area. We use the optical cross section measured from the photometry in Tab.~\ref{tab:cross}, which was derived assuming \(p_r=0.04\), and convert it to an equivalent icy-grain cross section by scaling
\begin{equation}
    C_{\rm ice}=C_{\rm e}\frac{p_r}{p_{\rm ice}}.
\end{equation}
For spherical grains, the corresponding water production rate is
\begin{equation}
    Q_{\rm H_2O,grain}
    =
    \frac{
    4\rho_{\rm ice} C_{\rm ice}
    \left|da/dt\right|
    }{
    m_{\rm H_2O}
    },
    \label{eq:q_from_cice}
\end{equation}
where \(\rho_{\rm ice}\) is the bulk density of water ice and \(m_{\rm H_2O}\) is the mass of one water molecule. In applying this equation, \(C_{\rm ice}\) and \(\left|da/dt\right|\) are converted to SI units, and \(\rho_{\rm ice}=0.92~{\rm g~cm^{-3}}\) is adopted. The resulting water-production rates are compared with published measurements in Fig.~\ref{Water_Production_Rates}.

\bibliography{sample701}{}
\bibliographystyle{aasjournalv7}



\end{document}